\RequirePackage{fix-cm}
\documentclass[smallextended]{svjour3}       
\smartqed  
\usepackage{graphicx}
\usepackage{color}
\usepackage{amsmath}
\usepackage{amssymb}
\begin{document}

\title{Constitutive equations for an electro-active polymer
}


\author{Mireille Tixier         \and
        Jo\"{e}l Pouget 
}


\institute{M. Tixier \at
              D\'{e}partement de Sciences Physiques, U.F.R. des Sciences, Universit\'{e} de Versailles Saint Quentin, 45, avenue des Etats-Unis, F-78035 Versailles \\
              Tel.: +33-139254519\\
              Fax: +33-139254523\\
              \email{mireille.tixier@uvsq.fr}           
           \and
           J. Pouget \at
              Sorbonne Universit\'{e}s, UPMC Univ. Paris 06, UMR 7190, Institut Jean le Rond d'Alembert, F-75005 Paris, France \\
              CNRS, UMR 7190,  Institut Jean le Rond d'Alembert, F-75005 Paris, France \\
              Tel.: +33-144275465\\
              \email{pouget@lmm.jussieu.fr}           
}

\date{Received: date / Accepted: date}

\maketitle

\begin{abstract}
Ionic electro-active polymers (E.A.P.) can be used as sensors or actuators.
For this purpose, a thin film of polyelectrolyte is saturated with a solvent
and sandwiched between two platinum electrodes. The solvent causes a
complete dissociation of the polymer and the release of small cations. The
application of an electric field across the thickness results in the bending
of the strip and vice versa.
The material is modelled by a two-phase continuous medium. The solid phase,
constituted by the polymer backbone inlaid with anions, is depicted as a
deformable porous media. The liquid phase is composed of the free cations
and the solvent (usually water). We used a coarse grain model.
The conservation laws of this system have been established in a previous
work. The entropy balance law and the thermodynamic relations are first
written for each phase, then for the complete material using a statistical
average technique and the material derivative concept. One deduces the
entropy production. Identifying generalized forces and fluxes provides the
constitutive equations of the whole system : the stress-strain relations
which satisfy a Kelvin-Voigt model, generalized Fourier's and Darcy's laws
and the Nernst-Planck equation.

\end{abstract}


\section{Introduction}
\label{intro} In a previous work presented by the authors \cite{Tixier}, 
conservation laws for an electro-active polymer have been established and discussed.
Especially, the equations of mass conservation, of the electric charge
conservation, the conservation of the momentum and different energy balance
equations at the macroscale of the material have been deduced using an
average technique for the different phases (solid and liquid). The present
work attempts the construction of constitutive equations. The latter are
deduced from the entropy balance law and thermodynamic relations. The
interest of such a formulation is that we arrive at tensorial and vectorial
constitutive equations for the macroscopic quantities for which the
constitutive coefficients can be expressed in terms of the microscopic
components of the electro-active polymers. \newline

\noindent More precisely, electro-active polymers can be classified in two
essential categories depending on their process of activation. The first
class is the electronic EAP and their actuation uses the electromechanical
coupling (linear or non linear coupling). These polymers are very similar to
piezoelectric materials. Their main drawback is an actuation which requires
very high voltage. The second class of EAP are the ionic polymers. They are
based on the ion transport due to applied electric voltage. This kind of EAP
exhibits very large transformation (large deflexion) in the presence of low
applied voltage (few volts). Their main drawback is that they operate best
in a humid environment and they must be encapsulated to operate in ambient
environment. \newline

\noindent The present paper places the emphasis on the ionic polymer metal
composite (IPMC). Such class of electro-active polymers is an active
material consisting in a thin membrane of polyelectrolyte (Nafion, for
instance) sandwiched on both sides with thin metal layers acting as
electrodes. The EAP can be deformed repetitively by applying a difference of
electric potential across the thickness of the material and it can quickly
recover its original configuration upon releasing the voltage. \newline
\noindent The mechanism of EAP deformation can be explained physically as
follows. Upon the application of an electric field across a moist polymer,
which is held between metallic electrodes attached across a partial section
of the EAP strip, bending of the EAP is produced (Fig. 1a). The positive counter ions
move towards the negative electrode (cathode), while negative ions that are
fixed to the polymer backbone experience an attractive force from positive
electrode (anode). At the same time water molecules in the EAP matrix
diffuse towards the region of the high positive ion concentration (near the
negative electrode) to equalize the charge distribution. As a result, the
region near the cathode swells and the region near the anode de-swells,
leading to stresses which cause the EAP strip to bend towards the positive
anode (Fig. 1b). When the electric field is released the EAP strip recover its initial
geometry. Conversely, a difference of electric potential is produced across
the EAP when it is suddenly bent. \newline

\begin{figure*}
 \includegraphics[width=0.45\textwidth]{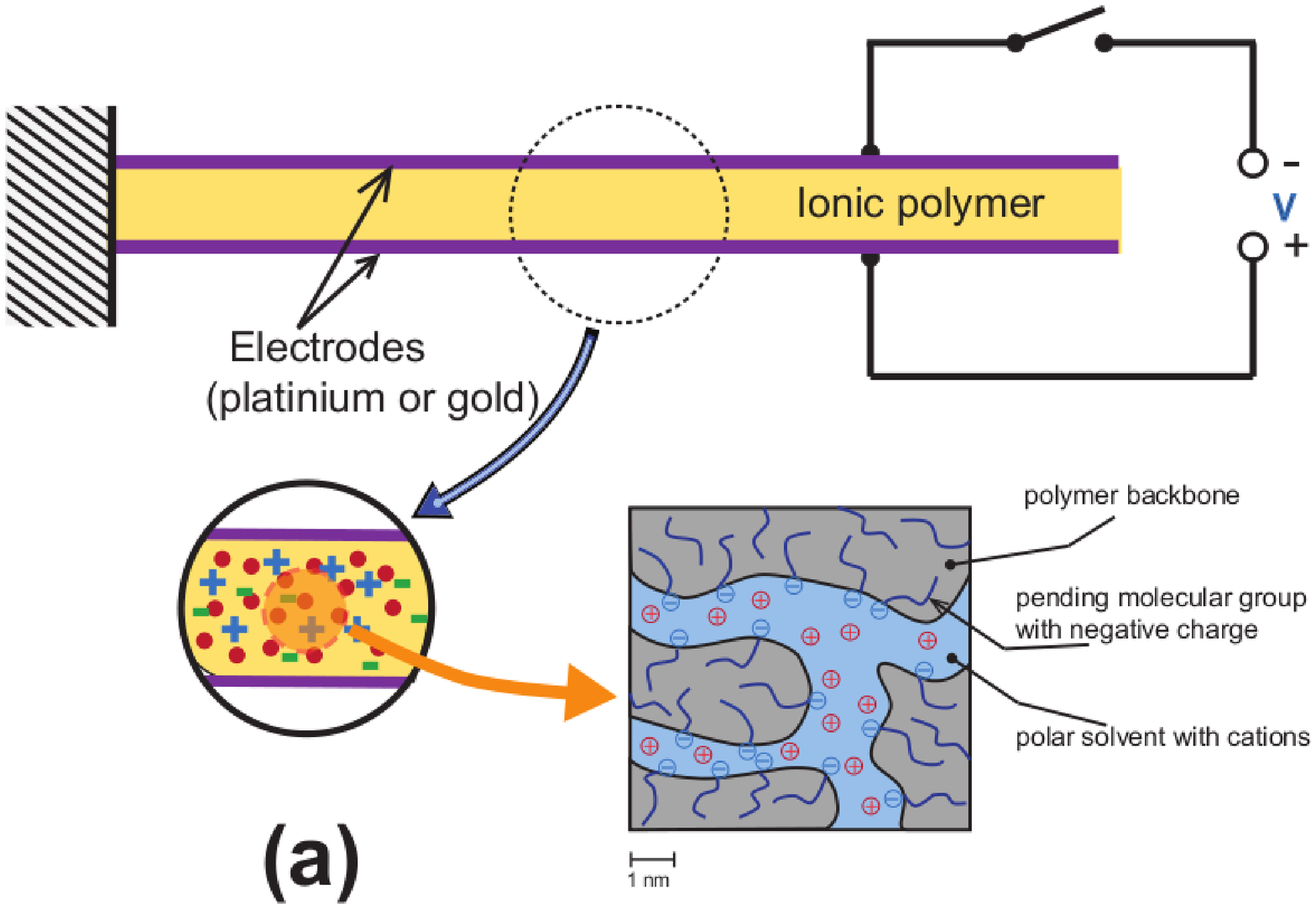}
 \qquad
 \includegraphics[width=0.45\textwidth]{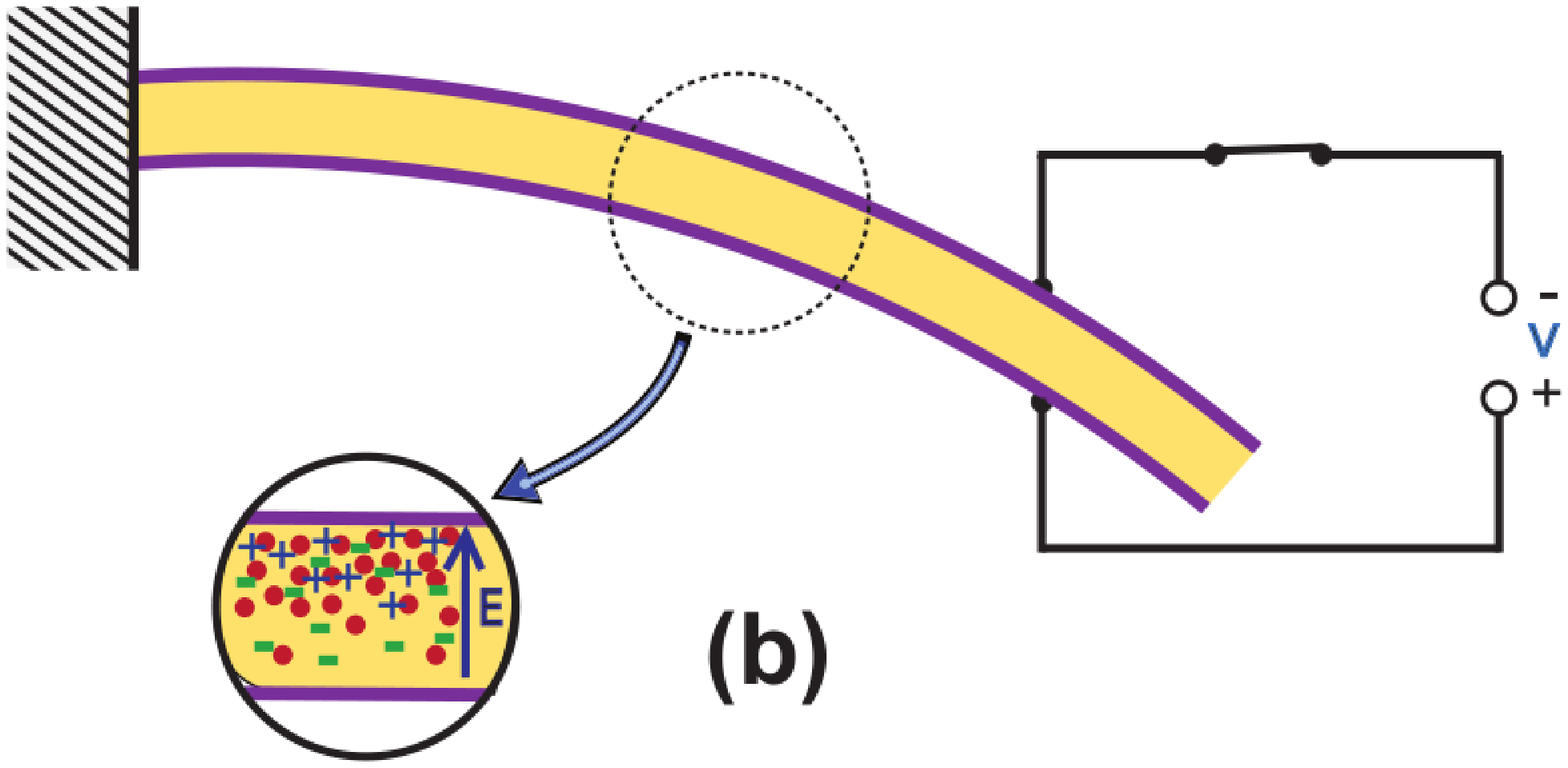}
\caption{Sketched representation of thin film of EAP : (a) undeformed strip. An inset 
gives the representative elementary volume (REV) containing the microscopic components 
of the polymer (Nafion)  and (b) the strip bending under an applied electric field. }
\label{fig:1}
\end{figure*}

\noindent Electromechanical coupling in ionic polymer membranes was
discovered over 50 years ago but has recently received renewed attention due
to the development of large strain actuators operating at low electric
fields. Modelling of EAP attracts scientists and engineers and a certain
number of approaches has been proposed to explain and quantify the physical
micro-mechanism which relate the EAP deformation to osmotic diffusion of
solvent and ions into the polymer. A micromechanical model has been developed
by Nemat-Nasser and Li \cite{nemat2000,nemat2002}. The model accounts for
the electromechanical and chemical-electric coupling of the ion transport,
electric field and elastic deformation to produce the response of the EAP.
The authors examine the field equations that place the osmotic stress in
evidence. They deduce a generalized Darcy's law and the balance law for the
ion flux - a kind of Nernst-Plank equation - deduced from the equation of
the electric charge conservation. A first simple macroscopic model was
proposed by DeGennes \textit{et al.} \cite{degennes}. The model describes
the coupling between the electric current density and the solvent (water)
flux. Shahinpoor \textit{et al.} \cite{shahinpoor1991,shahinpoor1994,shahinpoor1998}
report  the modelling of ion-exchange polymer-metal composites (IPMCs) based on 
an equation governing the ionic transport mechanism. The authors write  down the 
equations for the solvent concentration, the ionic concentration, and the relationship 
between stress, strain, electric field, heat flux and chemical  energy flux. The stress
tensor is related to the deformation gradient field by using a constitutive
equation of the neo-Hookean type. \newline

\noindent  The paper is divided in 8 Sections with 3 appendices. The next
section recalls the main results of the previous work \cite{Tixier} and
notations.  The section 3 concerns the entropy balance law at the
microscopic level and for the whole material over the R.V.E. (Representative Volume Element). 
The fundamental thermodynamic  relations are given in Section 4. The thermodynamic equations
are written for each phase (solid, solvent) and then for the complete
material leading to  Gibb's relation. On using the latter relation the
generalized forces and fluxes are identified in Section 5. The Section 6 is
devoted to constitutive equations,  especially the tensorial and vectorial
constitutive equations are deduced by invoking symmetry properties. A
detailed discussion on the results thus obtained  is presented in Section 7
and some estimates of the constitutive coefficients are given and compared
to the proposed approximations. The paper is closed with a brief  conclusion.

\section{Modelling and previous results}

\label{sec:1} The system we study is an ionic polymer-metal composite (IPMC); it consists
of a polyelectrolyte coated on both sides with thin metal layers acting as
electrodes. The electro-active polymer is saturated with water, which
results in a quasi-complete dissociation of the polymer : anions remain
bound to the polymer backbone, whereas small cations are released in water 
\cite{Futerko}. When an electric field perpendicular to the electrodes is
applied, the strip bends : cations are attracted by the negative electrode
and carry solvent away by osmosis. As a result, the polymer swells near the
negative electrode and contracts on the opposite side, leading to the bowing. \newline

\noindent The modelling of this system is detailled in our previous article \cite%
{Tixier}. The polymer chains are assimilated to a deformable porous medium
saturated by an ionic solution composed by water and cations. We suppose
that the solution is dilute. We depicted the complete material as the
superposition of three systems : a deformable solid made up of polymer
backbone negatively charged, a solvent (the water) and cations (see the inset of Fig. 1 for schematic representation). The three
components have different velocity fields, and the solid and liquid phases
are assumed to be incompressible phases separated by an interface whose
thickness is supposed to be negligible. We identify the quantities relative
to the different components \ by subscripts : $1$ refers to cations, $2$ to
solvent, $3$ to solid, $i$ to the interface and $4$ to the solution, that is
both components $1$ and $2$; the lack of subscript refers to the complete
material. Components $2$, $3$ and $4$ as well as the global material are
assimilated to continua. We venture the hypothesis that gravity and magnetic
field are negligible, so the only external force acting on the system is the
electric force. \newline

\noindent We describe this medium using a coarse-grained model developed for two-phase
mixtures \cite{Nigmatulin79,Nigmatulin90,Drew83,Drew98,Ishii06,Lhuillier03}. 
The microscopic scale is large enough to provide the continuum assumption, 
but small enough to enable the definition of a volume which contains a single 
phase ($3$ or $4$). At the macroscopic scale, we define a representative 
elementary volume (R.V.E.) which contains the two phases; it must be small 
enough so that average quantities relative to the whole material can be 
considered as local, and large enough so that this average is relevant.
A microscale Heaviside-like function of presence $\chi _{k}\left( 
\mathbf{r},t\right) $ has been defined for the phases $3$ and $4$%
\begin{equation}
\chi _{k}=1\;when\;phase\;k\;occupies\;point\;\mathbf{r}%
\;at\;time\;t,\quad \chi _{k}=0\;otherwise
\end{equation}%
The function of presence of the interface is the Dirac-like function $\chi
_{i}=-\boldsymbol{\nabla}\chi _{k}\cdot \mathbf{n_{k}}$ (in $%
m^{-1} $) where $\mathbf{n_{k}}$ is the outward-pointing unit normal to
the interface in the phase $k$. $\left\langle {}\right\rangle _{k}$ denotes
the average over the phase $k$ of a quantity relative to the phase $k$ only.
The macroscale quantities relative to the whole material are obtained by
statistically averaging the microscale quantities over the R.V.E., that is
by repeating many times the same experiment. We suppose that this average,
denoted by $\left\langle {}\right\rangle $, is equivalent to a volume
average (ergodic hypothesis) and commutes with the space and time
derivatives \cite{Drew83,Lhuillier03}. A macroscale quantity $g_{k}$
verifies%
\begin{equation}
g_{k}=\left\langle \chi _{k}g_{k}^{0}\right\rangle =\phi _{k}\left\langle
g_{k}^{0}\right\rangle _{k}
\end{equation}%
where $g_{k}^{0}$ is the corresponding microscale quantity and $\phi
_{k}=\left\langle \chi _{k}\right\rangle $ the volume fraction of the phase $%
k$. In the following, we use superscript $^{0}$ to indicate microscale
quantities; the macroscale quantities, which are averages defined all over
the material, are written without superscript. \newline

\noindent The conservation and balance laws of the polymer saturated with water have
been previously established \cite{Tixier}. To this end, we supposed that the
fluctuations of the following quantities are negligible on the R.V.E. scale
: the velocity $\mathbf{V_{k}}$ of each phase and interface, the
solid displacement vector $\mathbf{u_{3}}$, the cations molar
concentration $C$ and the electric field $\mathbf{E}$. Futhermore,
we admitted that the electric field is identical in all the phases and that
the solid and liquid phases are isotropic linear dielectrics. We thus
established the mass conservation law of the constituents and of the
complete material%
\begin{equation}
\begin{tabular}{l}
$\frac{\partial \rho _{k}}{\partial t}+div\left( \rho _{k}\mathbf{%
V_{k}}\right) =0\qquad for\;k=2,3$ \\ 
$\frac{\partial \rho }{\partial t}+div\left( \rho \mathbf{V}\right)
=0$%
\end{tabular}%
\end{equation}%
where $\rho _{k}=\phi _{k}\rho _{k}^{0}$ denotes the mass density of the
phase $k$ relative to the volume of the whole material. Maxwell's equations
and the constitutive relation can \ be written%
\begin{equation}
\mathbf{rot} \mathbf{E}=\mathbf{0}\qquad \qquad div%
\mathbf{D}=\rho Z\qquad \qquad \mathbf{D}=\varepsilon 
\mathbf{E}
\end{equation}%
with%
\begin{equation}
\varepsilon =\sum\limits_{k=3,4}\phi _{k}\varepsilon _{k}^{0}
\end{equation}%
where $\mathbf{D}$ is the electric displacement field, $Z$ the total
electric charge per unit of mass and $\varepsilon $ the permittivity. The
linear momentum and internal energy balance laws are 
\begin{equation}
\rho \frac{D\mathbf{V}}{Dt}= \mathbf{div} \boldsymbol{\sigma } +\rho Z\mathbf{E}
\end{equation}%
\begin{equation}
\rho \frac{D}{Dt}\left( \frac{U_{\Sigma }}{\rho }\right)
=\sum\limits_{k=3,4}\left( \boldsymbol{\sigma _{k}}:%
\boldsymbol{\nabla}\mathbf{V_{k}}\right) +%
\mathbf{i}\cdot\mathbf{E}-div\mathbf{Q}
\end{equation}%
where $\boldsymbol{\sigma }$ denotes the stress tensor, 
$\mathbf{i}$ the diffusion current, $\mathbf{Q}$ the heat
flux and $U_{\Sigma }$ the sum of the volume internal energies of the
different phases. These two relations use the material derivative $\frac{D}{%
Dt}$ defined in our previous paper \cite{Tixier} and reported in appendix A. \newline

\noindent The relative velocities of the different phases are negligible compared to
the velocities measured in the laboratory-frame. Let's take for example a
strip of Nafion which is $\textit{200}\;\mu m$ thick and $\textit{1.57}\;cm$ long, bending in
an electric field. The tip displacement is about $\textit{4}\;mm$ and it is obtained
in $\textit{1}\;s$ \cite{nemat2000}. The different phases velocities in the
laboratory-frame $\left\vert \mathbf{V_{k}^{0}}\right\vert $ are
close to $\textit{4}\;\textit{10}^{-3}\;m\;s^{-1}$ and the relative velocities $\left\vert%
\mathbf{V_{k}^{0}}-\mathbf{V}\right\vert $ to $%
\textit{2}\;\textit{10}^{-4}\;m\;s^{-1}$. So we can reasonably suppose that%
\begin{equation}
\left\vert \mathbf{V_{k}^{0}}-\mathbf{V}\right\vert
<<\left\vert \mathbf{V_{k}^{0}}\right\vert
\end{equation}%
The kinetic energy of the whole material is defined either as the sum $E_{c\Sigma }$ of the
kinetic energies of the constituents, or as the kinetic energy of the center
of mass of the constituents $E_{c}$ \cite{Tixier}. The difference between these two
quantities is%
\begin{equation}
E_{c\Sigma }-E_{c}=\frac{\rho _{3}\rho _{4}}{\rho }\left( \mathbf{%
V_{3}}-\mathbf{V_{4}}\right) ^{2}
\end{equation}%
and is negligible compared to the kinetic energies of each phases. On a
first approximation, we can therefore swap together $E_{c\Sigma }$ with $%
E_{c}$, and consequently $U_{\Sigma }$ with the internal energy of the whole system $U$.
Considering this hypothesis, the internal energy balance equation can be written%
\begin{equation}
\rho \frac{d}{dt}\left( \frac{U}{\rho }\right) =\boldsymbol{\sigma }:\boldsymbol{\nabla}%
\mathbf{V}+\mathbf{i^{\prime }}\cdot\mathbf{E}-div%
\mathbf{Q^{\prime }}  \label{U}
\end{equation}%
with%
\begin{equation}
\begin{tabular}{l}
$\mathbf{i^{\prime }}=\mathbf{I}-\rho Z\mathbf{V}
\simeq \rho _{1}Z_{1}\left( \mathbf{V_{1}}-\mathbf{V_{4}}
\right) +\sum\limits_{k=3,4}\rho _{k}Z_{k}\left( \mathbf{V_{k}}- 
\mathbf{V}\right) $ \\ 
$\mathbf{Q^{\prime }}=\mathbf{Q}-\sum\limits_{k=3,4}U_{k}
\left( \mathbf{V}-\mathbf{V_{k}}\right)
-\sum\limits_{k=3,4} \boldsymbol{\sigma _{k}}\cdot\left( 
\mathbf{V_{k}}- \mathbf{V}\right) $%
\end{tabular}%
\end{equation}
where $\mathbf{I}$ denotes the current density vector and
\begin{equation}
Z_{k}=Z_{k}^{0}\qquad for\;k=1,2,3\qquad \ and\qquad Z_{4}=\frac{\rho _{1}}{%
\rho _{4}}Z_{1}
\end{equation}%
To describe the systeme, we finally have 14 independent scalar equations
using 29 scalar variables ($\rho _{k}$, $\mathbf{V_{k}}$ ($k=1,3$), $\rho $, $%
\mathbf{V}$, $\boldsymbol{\sigma }$, $Z$, $%
\mathbf{E}$, $U$, $\mathbf{Q^{\prime }}$ and $%
\mathbf{D}$). 15 scalar equations are missing to close the system :
the constitutive relations. We will now establish them in the form of three
vectorial relations and one tensorial relation relating second-rank
symmetric tensors.


\section{Entropy balance law}

\label{sec:2} The microscale entropy balance laws of the solid and liquid phases can be
written

\begin{equation}
\frac{\partial S_{k}^{0}}{\partial t}+divS_{k}^{0}\mathbf{V_{k}^{0}}%
=-div\mathbf{\Sigma _{k}^{0}}+s_{k}^{0}\qquad \qquad k=3,4
\label{entropie_30}
\end{equation}%
where $S_{k}^{0}$, $\mathbf{\Sigma _{k}^{0}}$ and $s_{k}^{0}$
denote, respectively, the entropy density, the entropy flux vector and the
rate of entropy production of the phase $k$. \newline

\noindent Averaging over the R.V.E. gives, considering the interface condition $%
\mathbf{V_{1}^{0}}=\mathbf{V_{2}^{0}}=\mathbf{%
V_{3}^{0}}=\mathbf{V_{4}^{0}}=\mathbf{V_{i}^{0}}$

\begin{equation}
\frac{\partial S_{k}}{\partial t}+div\left( S_{k}\mathbf{V_{k}}%
\right) =s_{k}-div\mathbf{\Sigma _{k}}-\left\langle \mathbf{%
\Sigma _{k}^{0}}\cdot\mathbf{n_{k}}\chi _{i}\right\rangle
\end{equation}%
in which the macroscale entropy density $S_{k}$, the entropy flux vector $%
\mathbf{\Sigma _{k}}$ and the rate of entropy production $s_{k}$ are
defined by%
\begin{equation}
S_{k}=\left\langle \chi _{k}S_{k}^{0}\right\rangle \qquad \qquad 
\mathbf{\Sigma _{k}}=\left\langle \chi _{k}\mathbf{\Sigma
_{k}^{0}}\right\rangle \qquad \qquad s_{k}=\left\langle \chi
_{k}s_{k}^{0}\right\rangle
\end{equation}%
One points out that the quantities $S_{k}$ and $s_{k}$ are relative to the
volume of the whole material.
For the interface we obtain (see appendix B)%
\begin{equation}
\frac{\partial S_{i}}{\partial t}+div\left( S_{i}\mathbf{V_{i}}%
\right) =\sum\limits_{k=3,4}\left\langle \chi _{i}\mathbf{\Sigma
_{k}^{0}}\cdot\mathbf{n_{k}}\right\rangle +s_{i}
\end{equation}

\noindent The entropy balance law of the whole material is%
\begin{equation}
\rho \frac{D}{Dt}\left( \frac{S}{\rho }\right) =s-div\mathbf{\Sigma }
\end{equation}%
where :%
\begin{equation}
S=\sum\limits_{k=3,4,i}S_{k}\qquad \qquad s=\sum\limits_{k=3,4,i}s_{k}\qquad
\qquad \mathbf{\Sigma }=\sum\limits_{k=3,4}\mathbf{\Sigma
_{k}}
\end{equation}%
are the entropy density, the rate of entropy production and the entropy flux
vector of the complete material, respectively. In the barycentric frame of
reference, we derive%
\begin{equation}
\rho \frac{d}{dt}\left( \frac{S}{\rho }\right) =s-div\mathbf{\Sigma
^{\prime }}  \label{BilanEntr}
\end{equation}%
with%
\begin{equation}
\mathbf{\Sigma ^{\prime }}=\mathbf{\Sigma }%
-\sum\limits_{k=3,4}S_{k}\left( \mathbf{V}-\mathbf{V_{k}}%
\right)
\end{equation}


\section{Fundamental thermodynamic relations}
\label{sec:3}

\subsection{Thermodynamic relations for the solide phase}

\label{sec:31} For a solid phase with one constituent, the Gibb's relation can be written 
\cite{DeGroot}%
\begin{equation}
\rho _{3}^{0}\frac{d_{3}^{0}}{dt}\left( \frac{U_{3}^{0}}{\rho _{3}^{0}}%
\right) =p_{3}^{0}\frac{1}{\rho _{3}^{0}}\frac{d_{3}^{0}\rho _{3}^{0}}{dt}+%
\boldsymbol{\sigma _{3}^{0e}}^{s}:\frac{d_{3}^{0}%
\boldsymbol{\varepsilon _{3}^{0}}^{s}}{dt}+\rho
_{3}^{0}T_{3}^{0}\frac{d_{3}^{0}}{dt}\left( \frac{S_{3}^{0}}{\rho _{3}^{0}}%
\right)   \label{Gibbs-30}
\end{equation}%
where $T_{3}^{0}$ is the absolute temperature, $\boldsymbol{\varepsilon _{3}^{0}}$ 
the strain tensor, $\boldsymbol{\sigma _{3}^{0e}}$ the equilibrium stress tensor, $%
\boldsymbol{\sigma _{3}^{0e}}^{s}$ and $\boldsymbol{\varepsilon _{3}^{0}}^{s}$ the stress 
and strain deviator tensors, and $\frac{d_{3}^{0}}{dt}$ the particule derivative following the
microscale motion of the solid (see appendix A).  $p_{3}^{0}$ is the
pressure or negative one-third the trace of the microscopic equilibrium
stress tensor 
\begin{equation}
p_{3}^{0}=-\frac{1}{3}tr\left( \boldsymbol{\sigma_{3}^{0e}}\right) 
\end{equation}
Equation (\ref{Gibbs-30}) can also be deduced from equation (\ref{entropie_30})
and internal energy balance equation of the solid phase developped in \cite%
{Tixier} : indeed, the Gibbs relation is satisfied at equilibrium, so the
heat flux $\mathbf{Q_{3}^{0}}$, the diffusion current $%
\mathbf{i_{3}^{0}}$ and the rate of entropy production $s_{3}^{0}$
cancel; deformations are small and the stress tensor $\boldsymbol{\sigma _{3}^{0}}$ 
is equal to the equilibrium stress tensor $\boldsymbol{\sigma _{3}^{0e}}$. In addition, 
the solid phase is a closed system, consequently%
\begin{equation}
\mathbf{\Sigma _{3}^{0}}=\frac{\mathbf{Q_{3}^{0}}}{T_{3}^{0}}
\end{equation}
At the microscopic scale, Euler's homogeneous function theorem provides for the solid phase%
\begin{equation}
p_{3}^{0}=T_{3}^{0}S_{3}^{0}-U_{3}^{0}+\mu _{3}^{0}\rho _{3}^{0}
\end{equation}%
where $\mu _{3}^{0}$ denotes the chemical potential per unit of mass of the
solid constituent. As a result, Gibbs equation can be written%
\begin{equation}
\frac{d_{3}^{0}U_{3}^{0}}{dt}=T_{3}^{0}\frac{d_{3}^{0}S_{3}^{0}}{dt}+\mu%
_{3}^{0}\frac{d_{3}^{0}\rho _{3}^{0}}{dt}+\boldsymbol{\sigma _{3}^{0e}}^{s}:%
\frac{d_{3}^{0}}{dt}\boldsymbol{\varepsilon _{3}^{0}}^{s}
\end{equation}%
Differentiating Euler's relation and combining it with the Gibbs relation
leads to Gibbs-Duhem equation%
\begin{equation}
\frac{d_{3}^{0}p_{3}^{0}}{dt}=S_{3}^{0}\frac{d_{3}^{0}T_{3}^{0}}{dt}-%
\boldsymbol{\sigma _{3}^{0e}}^{s}:\frac{d_{3}^{0}}{dt} \boldsymbol{\varepsilon _{3}^{0}}^{s}%
+\rho _{3}^{0} \frac{d_{3}^{0}\mu _{3}^{0}}{dt}
\end{equation}
Let us assume that the fluctuations over the R.V.E. of the intensive
thermodynamical quantities $T_{3}^{0}$, $\mu _{3}^{0}$, $p_{3}^{0}$, the
displacement $\mathbf{u_{3}^{0}}$ and the equilibrium stress tensor $%
\boldsymbol{\sigma _{3}^{0e}}$ are negligible.
Supposing that the solid deformations are small, we obtain%
\begin{equation}
T_{3}=T_{3}^{0}\qquad \qquad \mu _{3}=\mu _{3}^{0}
\end{equation}%
\begin{equation}
\mathbf{u_{3}}=\mathbf{u_{3}^{0}}\qquad \qquad 
\boldsymbol{\varepsilon _{3}}=\boldsymbol{\varepsilon _{3}^{0}}=\frac{1}{2}\left(\boldsymbol{\nabla}%
\mathbf{u_{3}}+\boldsymbol{\nabla}\mathbf{u_{3}}^{T}\right)
\end{equation}%
\begin{equation}
p_{3}=p_{3}^{0}=-\frac{1}{3\phi _{3}}tr\boldsymbol{\sigma _{3}^{e}}\qquad \qquad
\boldsymbol{\sigma_{3}^{e}}=\phi _{3}\boldsymbol{\sigma _{3}^{0e}}=%
\boldsymbol{\sigma _{3}^{e}}^{s}-\phi _{3}p_{3} \boldsymbol{1}
\end{equation}%
where $\boldsymbol{1}$ denotes the second-rank identity
tensor and $\boldsymbol{\sigma _{3}^{e}}^{s}$ the
deviator part of $\boldsymbol{\sigma _{3}^{e}}$.
Considering the small deformation hypothesis, one easily derives (cf
appendix C)%
\begin{equation}
\begin{tabular}{ll}
$\frac{d_{3}U_{3}}{dt}=T_{3}\frac{d_{3}S_{3}}{dt}+\mu _{3}\frac{d_{3}\rho
_{3}}{dt}-p_{3}\frac{d_{3}\phi _{3}}{dt}+\boldsymbol{\sigma _{3}^{e}}^{s}:%
\frac{d_{3}}{dt}\boldsymbol{\varepsilon _{3}}^{s}$ & Gibbs \\ 
$\phi _{3}p_{3}=T_{3}S_{3}-U_{3}+\mu _{3}\rho _{3}$ & Euler \\ 
$\phi _{3}\frac{d_{3}p_{3}}{dt}=S_{3}\frac{d_{3}T_{3}}{dt}+\rho _{3}\frac{%
d_{3}\mu _{3}}{dt}-\boldsymbol{\sigma _{3}^{e}}^{s}:%
\frac{d_{3}}{dt}\boldsymbol{\varepsilon _{3}}^{s}$ & 
Gibbs-Duhem%
\end{tabular}%
\end{equation}

\subsection{Thermodynamic relations for the liquid phase}

\label{sec:32} According to S.R. De Groot and P. Mazur \cite{DeGroot}, the Gibbs relation
of a two-constituent fluid can be written as 
\begin{equation}
\frac{d_{4}^{0}}{dt}\left( \frac{U_{4}^{0}}{\rho _{4}^{0}}\right) =T_{4}^{0}%
\frac{d_{4}^{0}}{dt}\left( \frac{S_{4}^{0}}{\rho _{4}^{0}}\right) -p_{4}^{0}%
\frac{d_{4}^{0}}{dt}\left( \frac{1}{\rho _{4}^{0}}\right)
+\sum\limits_{k=1,2}\mu _{k}^{0}\frac{d_{4}^{0}}{dt}\left( \frac{\rho
_{k}^{\prime }}{\rho _{4}^{0}}\right) 
\end{equation}%
where $p_{4}^{0}$ is the fluid phase pressure, $\mu _{k}^{0}$ the mass
chemical potential of constituent $k$, and $\frac{d_{4}^{0}}{dt}$ the
particule derivative following the microscale motion of the liquid phase. $%
\rho _{k}^{\prime }$ are the mass densities of cations and solvent relative
to the solution volume%
\begin{equation}
\begin{tabular}{lll}
$\frac{\rho _{k}^{\prime }}{\rho _{4}^{0}}=\frac{\rho _{k}}{\rho _{4}}\qquad 
$ & $\rho _{1}^{\prime }=CM_{1}\qquad $ & $\rho _{2}^{\prime }=\frac{\rho
_{2}^{0}\phi _{2}}{\phi _{4}}$%
\end{tabular}%
\end{equation}%
$M_{1}$ denotes the cations molar mass and $C$ the cations molar
concentration relative to the liquid phase. As for the solid phase, one can
find out this equation combining the internal energy and entropy balance
laws and taking the limit at the equilibrium. Euler's
homogeneous function theorem takes on the following form at the microscopic scale%
\begin{equation}
U_{4}^{0}-T_{4}^{0}S_{4}^{0}+p_{4}^{0}=\sum\limits_{k=1,2}\mu _{k}^{0}\rho
_{k}^{\prime }
\end{equation}%
so that%
\begin{equation}
\frac{d_{4}^{0}U_{4}^{0}}{dt}=T_{4}^{0}\frac{d_{4}^{0}S_{4}^{0}}{dt}%
+\sum\limits_{k=1,2}\mu _{k}^{0}\frac{d_{4}^{0}\rho _{k}^{\prime }}{dt}
\end{equation}%
The Gibbs-Duhem relation of the liquid phase derives from the Gibbs and
Euler's relations%
\begin{equation}
\sum\limits_{k=1,2}\rho _{k}^{\prime }\frac{d_{4}^{0}\mu _{k}^{0}}{dt}%
=-S_{4}^{0}\frac{d_{4}^{0}T_{4}^{0}}{dt}+\frac{d_{4}^{0}p_{4}^{0}}{dt}
\end{equation}
We assume that the fluctuations of the intensive thermodynamic quantities
are negligible%
\begin{equation}
T_{4}=T_{4}^{0}\qquad \qquad \mu _{k}=\mu _{k}^{0}\qquad \qquad
p_{4}=p_{4}^{0}
\end{equation}%
Averaging the previous equations over the R.V.E., we obtain%
\begin{equation}
\begin{tabular}{ll}
$T_{4}\frac{d_{4}S_{4}}{dt}=\frac{d_{4}U_{4}}{dt}+p_{4}\frac{d_{4}\phi _{4}}{%
dt}-\sum\limits_{k=1,2}\mu _{k}\frac{d_{4}\rho _{k}}{dt}$ & Gibbs \\ 
$\phi _{4}p_{4}=T_{4}S_{4}-U_{4}+\sum\limits_{k=1,2}\mu _{k}\rho _{k}$ & 
Euler \\ 
$\phi _{4}\frac{d_{4}p_{4}}{dt}=S_{4}\frac{d_{4}T_{4}}{dt}%
+\sum\limits_{k=1,2}\rho _{k}\frac{d_{4}\mu _{k}}{dt}$ & Gibbs-Duhem%
\end{tabular}%
\end{equation}

\subsection{Thermodynamic relations for the complete material}

\label{sec:33} In order to write the thermodynamic relations of the complete material, we
make the hypothesis of local thermodynamic equilibrium; this requires among
other things that the heat diffuses well enough in the solid and the
solution so that temperature equilibrium is reached on the R.V.E.. We thus
can write%
\begin{equation}
\begin{tabular}{l}
$p=p_{3}=p_{4}$ \\ 
$T=T_{3}=T_{4}=T_{i}$%
\end{tabular}%
\end{equation}%
Otherwise, we have pointed out that the sum $U_{\Sigma }$ of the internal
energies of the constituents is close to the internal energy of the system $%
U $. Adding the Euler's relations of the solid and the liquid phases and the
interface, we thus obtain the Euler's relation of the whole material%
\begin{equation}
p=TS-U+\sum\limits_{k=1,2,3}\mu _{k}\rho _{k}
\end{equation}%
The Gibbs relation of the complete material is also obtained by addition%
\begin{equation}
T\frac{D}{Dt}\left( \frac{S}{\rho }\right) =\frac{D}{Dt}\left( \frac{U}{\rho 
}\right) +p\frac{D}{Dt}\left( \frac{1}{\rho }\right) -\frac{1}{\rho }%
\boldsymbol{\sigma _{3}^{e}}^{s}:\frac{d_{3} \boldsymbol{\varepsilon _{3}}^{s}}{dt}%
-\sum\limits_{1,2}\mu _{k}\frac{\rho _{4}}{\rho }\frac{d_{4}}{dt}\left( 
\frac{\rho _{k}}{\rho _{4}}\right)
\end{equation}%
The material derivative enables to follow the barycenters of each phase
during the motion. The solid phase is then supposed to be a closed system;
for this reason, no mass exchange term for the solid appears in this
relation. On the contrary, the solvent and the cations move at different
velocities; thence there is a mass exchange term concerning these two
constituents in the barycentric reference frame of the solution. The mass
exchanges of the three constituents appear if the particle derivative
following the motion of the whole material barycenter is used%
\begin{equation}
T\frac{d}{dt}\left( \frac{S}{\rho }\right) =\frac{d}{dt}\left( \frac{U}{\rho 
}\right) +p\frac{d}{dt}\left( \frac{1}{\rho }\right) -\sum\limits_{k=1,2,3}\mu
_{k}\frac{d}{dt}\left( \frac{\rho _{k}}{\rho }\right) -\frac{1}{\rho }%
\boldsymbol{\sigma _{3}^{e}}^{s}:\frac{d\boldsymbol{\varepsilon _{3}}^{s}}{dt}
\end{equation}%
This relation can also be obtained using the Gibbs relations of the
constituents; at equilibrium, indeed, the velocities of the two phases and
the interface are identical, in such a way that the particle derivatives are
the same%
\begin{equation}
\frac{d_{3}}{dt}\equiv \frac{d_{4}}{dt}\equiv \frac{d_{i}}{dt}\equiv \frac{d%
}{dt}
\end{equation}%
A third approach is to combine the balance equations of internal energy and
entropy of the complete material, and to take the limit at equilibrium.\newline

\noindent Considering the small deformations hypothesis and neglecting the relative
velocities compared to the velocities in the laboratory-frame, we derive%
\begin{equation}
\boldsymbol{\sigma _{3}^{e}}^{s}:\boldsymbol{\nabla}\mathbf{V_{3}}=\boldsymbol{%
\sigma _{3}^{e}}^{s}:\frac{d_{3}\boldsymbol{\varepsilon_{3}}^{s}}{dt}\simeq %
\boldsymbol{\sigma _{3}^{e}}^{s}: \frac{d\boldsymbol{\varepsilon _{3}}^{s}}{dt}\simeq %
\boldsymbol{\sigma _{3}^{e}}^{s}:\boldsymbol{\nabla}\mathbf{V}
\end{equation}%
The equilibrium stress tensor of the complete material is written as follows 
\begin{equation}
\boldsymbol{\sigma ^{e}}=\boldsymbol{\sigma _{3}^{e}}+\boldsymbol{\sigma_{4}^{e}}=%
-p\boldsymbol{1}+\boldsymbol{\sigma _{3}^{e}}^{s}  \label{ContEquil}
\end{equation}%
with%
\begin{equation}
\boldsymbol{\sigma _{4}^{e}}=-\phi _{4}p_{4} \boldsymbol{1}
\end{equation}%
As a result%
\begin{equation}
\boldsymbol{\sigma ^{e}}^{s}=\boldsymbol{\sigma _{3}^{e}}^{s}
\end{equation}%
Finally, Gibbs relation is%
\begin{equation}
T\frac{d}{dt}\left( \frac{S}{\rho }\right) =\frac{d}{dt}\left( \frac{U}{\rho 
}\right) +p\frac{d}{dt}\left( \frac{1}{\rho }\right)
-\sum\limits_{k=1,2,3}\mu _{k}\frac{d}{dt}\left( \frac{\rho _{k}}{\rho }%
\right) -\frac{1}{\rho }\boldsymbol{\sigma ^{e}}^{s}:%
\boldsymbol{\nabla}\mathbf{V}  \label{Gibbs}
\end{equation}
Differentiating Euler's relation and combining it with Gibbs relation, the
Gibbs-Duhem relation takes on the form 
\begin{equation}
\frac{dp}{dt}=S\frac{dT}{dt}+\sum\limits_{k=1,2,3}\rho _{k}\frac{d\mu _{k}}{%
dt}-\boldsymbol{\sigma ^{e}}^{s}:\boldsymbol{\nabla}\mathbf{V}
\end{equation}


\section{Generalized forces and fluxes}
\label{sec:4}

\subsection{Entropy production}

\label{sec:41} The stress tensor is composed of two parts : the equilibrium stress tensor $%
\boldsymbol{\sigma ^{e}}$ and the viscous stress tensor $\boldsymbol{\sigma ^{v}}$, which vanishes at
equilibrium. Considering (\ref{ContEquil}), the complete stress tensor can
be written as 
\begin{equation}
\boldsymbol{\sigma }=\boldsymbol{\sigma _{3}}+\boldsymbol{\sigma _{4}}=\boldsymbol{\sigma ^{e}}+%
\boldsymbol{\sigma ^{v}}=-p\boldsymbol{1}+\boldsymbol{\sigma ^{e}}^{s}+\boldsymbol{\sigma ^{v}}
\end{equation}%
Combining the internal energy and entropy equations (\ref{U}) and (\ref%
{BilanEntr}) with the Gibbs relation (\ref{Gibbs}) yields%
\begin{equation}
\begin{tabular}{l}
$s-div\mathbf{\Sigma ^{\prime }}=\frac{1}{T}\boldsymbol{\sigma ^{v}}:\boldsymbol{\nabla}%
\mathbf{V}+\frac{1}{T}\mathbf{i^{\prime }}\cdot\mathbf{E}%
-\frac{1}{T^{2}}\mathbf{Q^{\prime }}\cdot\boldsymbol{\nabla} T$ \\ 
$\qquad +\sum\limits_{1,2,3}\rho _{k}\left( \mathbf{V}-%
\mathbf{V_{k}}\right) \cdot\boldsymbol{\nabla} \frac{\mu _{k}}{T}%
 -div\left[ \frac{\mathbf{Q^{\prime }}}{T}+\sum%
\limits_{1,2,3}\frac{\mu _{k}\rho _{k}}{T}\left( \mathbf{V}-%
\mathbf{V_{k}}\right) \right] $%
\end{tabular}%
\end{equation}%
We can then identify the rate of entropy production $s$ and the entropy flux
vector $\mathbf{\Sigma ^{\prime }}$ 
\begin{equation}
\begin{tabular}{l}
$s=\frac{1}{T}\boldsymbol{\sigma ^{v}}:\boldsymbol{\nabla}\mathbf{V}+\frac{1}{T}%
\mathbf{i^{\prime }}\cdot\mathbf{E}-\frac{1}{T^{2}}\mathbf{Q^{\prime }}\cdot%
\boldsymbol{\nabla} T+\sum\limits_{k=1,2,3}\rho _{k}\left( \mathbf{V}%
-\mathbf{V_{k}}\right) \cdot\boldsymbol{\nabla}\frac{\mu _{k}}{T}$ \\ 
$\mathbf{\Sigma ^{\prime }}=\frac{\mathbf{Q^{\prime }}}{T}%
+\sum\limits_{k=1,2,3}\frac{\mu _{k}\rho _{k}}{T}\left( \mathbf{V}-%
\mathbf{V_{k}}\right) $%
\end{tabular}%
\end{equation}

\subsection{Identification of the generalized forces and fluxes}

\label{sec:42} A second rank tensor is the sum of three parts : a spherical tensor, a
deviator tensor (labeled with $^{s}$) and an antisymmetric tensor (labeled
with $^{a}$)%
\begin{equation}
\boldsymbol{\nabla}\mathbf{V}=\frac{1}{3}\left(div\mathbf{V}\right)%
\boldsymbol{1}+\boldsymbol{\nabla}\mathbf{V}^{s}+\boldsymbol{\nabla}\mathbf{V}^{a}
\end{equation}%
where%
\begin{equation}
\boldsymbol{\nabla}\mathbf{V}^{s}=\frac{1}{2} \left( \boldsymbol{\nabla}\mathbf{V}+%
\boldsymbol{\nabla}\mathbf{V}^{T}\right) -\frac{1}{3}\left( div\mathbf{V}%
\right) \boldsymbol{1}
\end{equation}%
The viscous stress tensor is symmetric, so%
\begin{equation}
\boldsymbol{\sigma ^{v}}=\frac{1}{3}tr\left( 
\boldsymbol{\sigma ^{v}}\right) \boldsymbol{1}+\boldsymbol{\sigma ^{v}}^{s}
\end{equation}
In the entropy production $s$ appear the three mass diffusion fluxes
relative to the barycentric reference frame $\rho _{k}\left( \mathbf{%
V_{k}}-\mathbf{V}\right) $ with $k=1,2,3$. The sum of these three
fluxes is zero, so only two of them are linearly independant. We define the
following equivalent fluxes%
\begin{equation}
\mathbf{J_{1}}=\rho _{1}\left( \mathbf{V_{1}}-%
\mathbf{V_{2}}\right) \qquad \qquad \qquad \mathbf{J_{4}}%
=\rho _{4}\left( \mathbf{V_{4}}-\mathbf{V_{3}}\right)
\end{equation}%
which are respectively the mass diffusion flux of the cations in the
solution and the mass diffusion flux of the solution in the solid. These two
fluxes are linearly independant. The diffusion current $\mathbf{%
i^{\prime }}$ and the fluxes $\rho _{k}\left( \mathbf{V_{k}}-%
\mathbf{V}\right) $ can be expressed as functions of $%
\mathbf{J_{1}}$ and $\mathbf{J_{4}}$, then the entropy
production takes on the following form 
\begin{equation}
\begin{tabular}{ll}
$s$ & $=\frac{1}{3T}tr\left( \boldsymbol{\sigma ^{v}} \right) \cdot div\mathbf{V}$ \\ 
& $+\mathbf{Q^{\prime }}\cdot \boldsymbol{\nabla}\frac{1}{T}+\frac{\rho
_{2}}{\rho _{4}}\left[ \frac{1}{T}Z_{1}\mathbf{E}-\boldsymbol{\nabla}\frac{\mu _{1}}{T}+ %
\boldsymbol{\nabla}\frac{\mu _{2}}{T}\right] \cdot%
\mathbf{J_{1}}$ \\ 
& $+\frac{\rho _{3}}{\rho }\left[ \frac{1}{T}\left( \frac{\rho _{1}}{\rho
_{4}}Z_{1}-Z_{3}\right) \mathbf{E}-\frac{\rho _{1}}{\rho _{4}}%
\boldsymbol{\nabla}\frac{\mu _{1}}{T}-\frac{\rho _{2}}{\rho _{4}}%
\boldsymbol{\nabla}\frac{\mu _{2}}{T}+\boldsymbol{\nabla}\frac{\mu _{3}}{%
T}\right] \cdot \mathbf{J_{4}}$ \\ 
& $+\frac{1}{T}\boldsymbol{\sigma ^{v}}^{s}: \boldsymbol{\nabla}\mathbf{V}^{s}$%
\end{tabular}%
\end{equation}%
This expression places in evidence one scalar flux $\frac{1}{3}tr\left( 
\boldsymbol{\sigma ^{v}}\right) $, three vectorial
fluxes $\mathbf{Q^{\prime }}$, $\mathbf{J_{1}}$, $%
\mathbf{J_{4}}$ and one second-rank tensorial flux $\boldsymbol{\sigma ^{v}}^{s}$ 
along with the associated generalized forces%
\begin{equation}
\begin{tabular}{|l|l|}
\hline
Fluxes & Forces \\ \hline
$\frac{1}{3}tr \boldsymbol{\sigma ^{v}} $
& $\frac{1}{T}div\mathbf{V}$ \\ \hline
$\mathbf{Q^{\prime }}$ & $\boldsymbol{\nabla}\frac{1}{T}$ \\ \hline
$\mathbf{J_{1}}$ & $\frac{\rho _{2}}{\rho _{4}}\left[ \frac{1}{T}%
Z_{1}\mathbf{E}-\boldsymbol{\nabla}\frac{\mu _{1}}{T}+%
\boldsymbol{\nabla}\frac{\mu _{2}}{T}\right] $ \\ \hline
$\mathbf{J_{4}}$ & $\frac{\rho _{3}}{\rho }\left[ \frac{1}{T}\left( 
\frac{\rho _{1}}{\rho _{4}}Z_{1}-Z_{3}\right) \mathbf{E}-\frac{\rho
_{1}}{\rho _{4}}\boldsymbol{\nabla}\frac{\mu _{1}}{T}-\frac{\rho _{2}}{%
\rho _{4}}\boldsymbol{\nabla}\frac{\mu _{2}}{T}+\boldsymbol{\nabla}\frac{%
\mu _{3}}{T}\right] $ \\ \hline
$\boldsymbol{\sigma ^{v}}^{s}$ & $\frac{1}{T} \boldsymbol{\nabla}\mathbf{V}^{s}$ \\ \hline
\end{tabular}
\label{IdentFlux}
\end{equation}


\section{Constitutive equations}
\label{sec:5}

\subsection{Tensorial constitutive equation}

\label{sec:51} We assume that the medium is isotropic. According to Curie dissymmetry
principle, there can not be any coupling between fluxes and forces whose
tensorial ranks differs from one unit. Moreover, we suppose that coupling
between fluxes and different tensorial rank forces are negligible, which is
a generally accepted hypothesis \cite{DeGroot}. Consequently, the scalar
constitutive equation requires only one scalar phenomenological coefficient $%
L_{1}$%
\begin{equation}
\frac{1}{3}tr\left( \boldsymbol{\sigma ^{v}}\right) =\frac{L_{1}}{T}div\mathbf{V}
\end{equation}%
In the same way, the tensorial flux $\boldsymbol{\sigma^{v}}^{s}$ is related to the 
generalized force $\frac{1}{T}\boldsymbol{\nabla}\mathbf{V}^{s}$ by a fourth-rank tensorial
phenomenological coefficient $\boldsymbol{L_{2}}$%
\begin{equation}
\boldsymbol{\sigma ^{v}}^{s}=\boldsymbol{L_{2}}\frac{1}{T} %
\mathbf{\boldsymbol{\nabla} V}^{s}
\end{equation}%
Because of the isotropy of the medium, tensor $\boldsymbol{L_{2}}$
is isotropic and requires only three scalar coefficients \cite{DeGroot}.
Furthermore, tensors $\boldsymbol{\sigma ^{v}}^{s}$ and $\boldsymbol{\nabla}\mathbf{V}^{s}$ 
are deviatoric, so%
\begin{equation}
\boldsymbol{\sigma ^{v}}^{s}=\frac{L_{2}}{T} \boldsymbol{\nabla}\mathbf{V}^{s}
\end{equation}%
where $L_{2}$ is a scalar coefficient. Setting out $L_{1}^{\prime }=L_{1}-%
\frac{L_{2}}{3}$, the viscous stress tensor is finally given by 
\begin{equation}
\boldsymbol{\sigma ^{v}}=\frac{L_{1}^{\prime }}{T} \left( div\mathbf{V}\right) 
\boldsymbol{1}+\frac{L_{2}}{2T}\left( \boldsymbol{\nabla}\mathbf{%
V}+\boldsymbol{\nabla}\mathbf{V}^{T}\right)
\end{equation}
Assuming that the complete material satisfies the Hooke's law at
equilibrium, the equilibrium stress tensor can be written as 
\begin{equation}
\boldsymbol{\sigma ^{e}}=\lambda \left( tr \boldsymbol{\varepsilon }\right) %
\boldsymbol{1}+2G\boldsymbol{\varepsilon }=\frac{3\lambda +2G}{3}\left( tr%
\boldsymbol{\varepsilon } \right) \boldsymbol{1}+2G\boldsymbol{\varepsilon }^{s}
\end{equation}%
where $\lambda $ and $G$ denote the first Lam\'{e} constant and the shear
modulus of the complete material, respectively, and where the material
strain is defined by%
\begin{equation}
\boldsymbol{\varepsilon }=\frac{1}{2}\left( \boldsymbol{\nabla}\mathbf{u}%
+\boldsymbol{\nabla}\mathbf{u}^{T} \right) \qquad or\qquad %
\overset{%
\bullet }{\boldsymbol{\varepsilon }}=\frac{1}{2}\left( %
\boldsymbol{\nabla}\mathbf{V}+\boldsymbol{\nabla}\mathbf{V}^{T}\right)
\end{equation}%
$\mathbf{u}$ is the displacement vector. Supposing that the fluid is
newtonian and stokesian, the pressure is%
\begin{equation}
p=-\frac{1}{3}tr\left( \boldsymbol{\sigma ^{e}}\right)
=\left( \lambda +\frac{2}{3}G\right) tr\boldsymbol{\varepsilon }
\end{equation}%
The stress tensor of the complete material thus satisfies a Kelvin-Voigt
model%
\begin{equation}
\boldsymbol{\sigma }=\lambda \left( tr\boldsymbol{\varepsilon }\right) \boldsymbol{1}+2G%
\boldsymbol{\varepsilon }+\frac{L_{1}^{\prime }}{T} \left( tr\overset{\bullet }%
{\boldsymbol{\varepsilon }} \right) \boldsymbol{1}+\frac{L_{2}}{T}\overset{\bullet }%
{\boldsymbol{\varepsilon }}  \label{LR}
\end{equation}

\subsection{Chemical potentials}

\label{sec:52} The liquid phase is a dilute solution of strong electrolyte. Molar chemical
potentials of the three constituents $\mu _{i,mol}$ can be written on a
first approximation \cite{Diu}%
\begin{equation}
\begin{tabular}{l}
$\mu _{1,mol}\left( T,p,x\right) =\mu _{1,mol}^{0}\left( T,p\right) +RT\ln
x+o(\sqrt{x})$ \\ 
$\mu _{2,mol}\left( T,p,x\right) =\mu _{2,mol}^{0}\left( T,p\right)
-RTx+o(x^{3/2})$ \\ 
$\mu _{3,mol}\left( T,p,x\right) =\mu _{3,mol}^{0}\left( T\right) $%
\end{tabular}%
\end{equation}%
where $R=8,314\;J\;K^{-1}$ is the gas constant and $x$ the molar fraction of
the cations in the solution%
\begin{equation}
x=C\frac{M_{2}}{\rho _{2}^{0}}
\end{equation}%
$\mu _{2,mol}^{0}$ and $\mu _{3,mol}^{0}$ denote the chemical potentials of
the single solid and solvent, and $\mu _{1,mol}^{0}$ depends on the solvent
and the solute. Mass chemical potentials $\mu _{i}=\frac{\mu _{i,mol}}{M_{i}}
$ can then be written%
\begin{equation}
\begin{tabular}{l}
$\mu _{1}\left( T,p,x\right) \simeq \mu _{1}^{0}\left( T,p\right) +\frac{RT}{M_{1}}%
\ln \left( C\frac{M_{2}}{\rho _{2}^{0}}\right) $ \\ 
$\mu _{2}\left( T,p,x\right) \simeq \mu _{2}^{0}\left( T,p\right) -\frac{RT}{\rho _{2}^{0}}C$ \\ 
$\mu _{3}\left( T,p,x\right) =\mu _{3}^{0}\left( T\right) $%
\end{tabular}%
\end{equation}%
Using the Gibbs-Duhem's relations for the solid and the liquid phases, we
obtain%
\begin{equation}
\begin{tabular}{l}
$\boldsymbol{\nabla}\mu _{1}=-\frac{S_{1}}{\rho _{1}}\boldsymbol{\nabla}%
T+\frac{v_{1}}{M_{1}}\boldsymbol{\nabla} p+\frac{RT\rho _{2}^{0}}{%
M_{2}M_{1}C}\boldsymbol{\nabla}\left( \frac{CM_{2}}{\rho _{2}^{0}}\right) $
\\ 
$\boldsymbol{\nabla}\mu _{2}=-\frac{S_{2}}{\rho _{2}}\boldsymbol{\nabla} T+\frac{v_{2}}{M_{2}} %
\boldsymbol{\nabla} p-\frac{RT}{M_{2}}\boldsymbol{\nabla}\left( \frac{CM_{2}}{\rho _{2}^{0}}\right) $ \\ 
$\boldsymbol{\nabla}\mu _{3}=-\frac{S_{3}}{\rho _{3}}\boldsymbol{\nabla}%
T $%
\end{tabular}%
\end{equation}%
where $v_{i}$ denotes the partial molar volume of the constituent $i$.

\subsection{Vectorial constitutive equations}

\label{sec:53} Vectorial constitutive equations require nine phenomenogical coefficients.
These coefficients are a priori second-rank tensors; considering the
isotropy of the medium, they can be replaced by scalars%
\begin{equation}
\begin{tabular}{l}
$\mathbf{Q^{\prime }}=L_{3}\boldsymbol{\nabla}\frac{1}{T}+L_{4}%
\frac{\rho _{2}}{\rho _{4}}\left[ \frac{1}{T}Z_{1}\mathbf{E}-%
\boldsymbol{\nabla} \frac{\mu _{1}}{T}+\boldsymbol{\nabla} \frac{\mu _{2}}{%
T}\right] $ \\ 
$\quad +L_{5}\frac{\rho _{3}}{\rho }\left[ \frac{1}{T}\left( \frac{\rho _{1}%
}{\rho _{4}}Z_{1}-Z_{3}\right) \mathbf{E}-\frac{\rho _{1}}{\rho _{4}}%
\boldsymbol{\nabla} \frac{\mu _{1}}{T}-\frac{\rho _{2}}{\rho _{4}}%
\boldsymbol{\nabla} \frac{\mu _{2}}{T}+\boldsymbol{\nabla} \frac{\mu _{3}}{%
T}\right] $%
\end{tabular}%
\end{equation}%
\begin{equation}
\begin{tabular}{l}
$\mathbf{J_{1}}=L_{4}\boldsymbol{\nabla} \frac{1}{T}+L_{7}\frac{%
\rho _{2}}{\rho _{4}}\left[ \frac{1}{T}Z_{1}\mathbf{E}-%
\boldsymbol{\nabla} \frac{\mu _{1}}{T}+\boldsymbol{\nabla} \frac{\mu _{2}}{%
T}\right] $ \\ 
$\quad +L_{8}\frac{\rho _{3}}{\rho }\left[ \frac{1}{T}\left( \frac{\rho _{1}%
}{\rho _{4}}Z_{1}-Z_{3}\right) \mathbf{E}-\frac{\rho _{1}}{\rho _{4}}%
\boldsymbol{\nabla} \frac{\mu _{1}}{T}-\frac{\rho _{2}}{\rho _{4}}%
\boldsymbol{\nabla} \frac{\mu _{2}}{T}+\boldsymbol{\nabla} \frac{\mu _{3}}{%
T}\right] $%
\end{tabular}%
\end{equation}%
\begin{equation}
\begin{tabular}{l}
$\mathbf{J_{4}}=L_{5}\boldsymbol{\nabla} \frac{1}{T}+L_{8}\frac{%
\rho _{2}}{\rho _{4}}\left[ \frac{1}{T}Z_{1}\mathbf{E}-%
\boldsymbol{\nabla} \frac{\mu _{1}}{T}+\boldsymbol{\nabla} \frac{\mu _{2}}{%
T}\right] $ \\ 
$\quad +L_{11}\frac{\rho _{3}}{\rho }\left[ \frac{1}{T}\left( \frac{\rho _{1}%
}{\rho _{4}}Z_{1}-Z_{3}\right) \mathbf{E}-\frac{\rho _{1}}{\rho _{4}}%
\boldsymbol{\nabla} \frac{\mu _{1}}{T}-\frac{\rho _{2}}{\rho _{4}}%
\boldsymbol{\nabla} \frac{\mu _{2}}{T}+\boldsymbol{\nabla} \frac{\mu _{3}}{%
T}\right] $%
\end{tabular}%
\end{equation}%
Onsager reciprocal relations lead otherwise to%
\begin{equation}
L_{6}=L_{4}\qquad \qquad L_{9}=L_{5}\qquad \qquad L_{10}=L_{8}
\end{equation}
Considering that the solution is dilute and using the expressions obtained
for the chemical potentials, the heat flux writes%
\begin{equation}
\mathbf{Q^{\prime }}=-\frac{A_{QT}}{T^{2}}\boldsymbol{\nabla} T+%
\frac{A_{QE}}{T}\mathbf{E}+\frac{A_{QP}}{T}\boldsymbol{\nabla} %
p+A_{QC}\boldsymbol{\nabla} C
\end{equation}%
with%
\begin{equation}
\begin{tabular}{l}
$A_{QT}\simeq L_{3}+L_{4}\left[ -\mu _{1}^{0}-\frac{RT}{M_{1}}\ln \left( 
\frac{CM_{2}}{\rho _{2}^{0}}\right) +\mu _{2}^{0}-\frac{RT}{\rho _{2}^{0}}C-%
\frac{S_{1}T}{\rho _{1}}+\frac{S_{2}T}{\rho _{2}}\right] $ \\ 
$\quad +\frac{\rho _{3}}{\rho }L_{5}\left[ -\frac{\rho _{1}}{\rho _{2}}%
\left( \mu _{1}^{0}+\frac{RT}{M_{1}}\ln \frac{CM_{2}}{\rho _{2}^{0}}%
\right) -\mu _{2}^{0}+\frac{RTC}{\rho _{2}^{0}}+\mu _{3}^{0}-\frac{%
S_{4}T}{\rho _{4}}+\frac{S_{3}T}{\rho _{3}}\right] $ \\
$A_{QE}\simeq L_{4}Z_{1}+\frac{\rho _{3}}{\rho \rho _{4}}\left( \rho
_{1}Z_{1}-\rho _{2}Z_{3}\right) L_{5}$ \\ 
$A_{QP}\simeq L_{4}\left( \frac{v_{2}}{M_{2}}-\frac{v_{1}}{M_{1}}\right) -%
\frac{\rho _{3}}{\rho \rho _{4}}\phi _{4}L_{5}$ \\ 
$A_{QC}\simeq -\frac{R}{M_{1}C}L_{4}$%
\end{tabular}%
\end{equation}%
According to the definition of the partial molar volumes, indeed%
\begin{equation}
\rho _{1}\frac{v_{1}}{_{M_{1}}}+\rho _{2}\frac{v_{2}}{_{M_{2}}}=\phi _{4}
\end{equation}%
Likewise, the mass diffusion flux of the cations in the solution can be
written as 
\begin{equation}
\mathbf{J_{1}}=-\frac{A_{1T}}{T^{2}}\boldsymbol{\nabla} T+\frac{%
A_{1E}}{T}\mathbf{E}+\frac{A_{1P}}{T}\boldsymbol{\nabla} p+A_{1C}%
\boldsymbol{\nabla} C
\end{equation}%
with%
\begin{equation}
\begin{tabular}{l}
$A_{1T}\simeq L_{4} +L_{7}\left( \mu _{2}^{0}-\mu _{1}^{0}-\frac{RT}{M_{1}} \ln%
\frac{CM_{2}}{\rho _{2}^{0}}-\frac{RTC}{\rho _{2}^{0}}+ T \frac{S_{2}%
}{\rho _{4}} -T \frac{S_{1}}{\rho _{1}} \right) $ \\
$+ \frac{\rho _{3} L_{8}}{\rho _{4}\rho }\left[ \frac{RT\rho _{1}}{%
M_{1}}\left( 1-\ln \frac {C M_{2}}{\rho _{2}^{0}}\right) -\rho _{1}\mu%
_{1}^{0}-\mu _{2}^{0}+\rho _{4}\mu _{3}^{0} - T S_{2} - T S_{1} %
-\frac{T \rho _{4} S_{3}}{\rho _{3}} \right]  $ \\  
$A_{1E}\simeq L_{7}Z_{1}+\frac{\rho _{3}}{\rho }L_{8}\left( \frac{\rho _{1}}{%
\rho _{4}}Z_{1}-Z_{3}\right) $ \\ 
$A_{1P}\simeq L_{7}\left( \frac{v_{2}}{M_{2}}-\frac{v_{1}}{M_{1}}\right) -%
\frac{\rho _{3}\phi _{4}}{\rho \rho _{4}}L_{8}$ \\ 
$A_{1C}\simeq -\frac{R}{M_{1}C}L_{7}$%
\end{tabular}%
\end{equation}%
and the mass diffusion flux of the solution in the solid is%
\begin{equation}
\mathbf{J_{4}}=-\frac{A_{4T}}{T^{2}}\boldsymbol{\nabla} T+\frac{%
A_{4E}}{T}\mathbf{E}+\frac{A_{4P}}{T}\boldsymbol{\nabla} p+A_{4C}%
\boldsymbol{\nabla} C
\end{equation}%
with%
\begin{equation}
\begin{tabular}{l}
$A_{4T}\simeq L_{5}+L_{8}\left[ \mu _{2}^{0}-\frac{RT}{\rho _{2}^{0}}C-\mu
_{1}^{0}-\frac{RT}{M_{1}}\ln \left( C\frac{M_{2}}{\rho _{2}^{0}}\right) -%
\frac{TS_{1}}{\rho _{1}}+\frac{TS_{2}}{\rho _{2}}\right] $ \\ 
$+\frac{\rho _{3}L_{11}}{\rho }\left[ \frac{RTC}{\rho _{2}^{0}}-\frac{%
RTC}{\rho _{4}^{0}}\ln \frac{C M_{2}}{\rho _{2}^{0}}-\frac{\rho _{1}\mu
_{1}^{0}}{\rho _{4}}-\mu _{2}^{0}+\mu _{3}^{0}-\frac{TS_{1}}{\rho _{4}}-%
\frac{TS_{2}}{\rho _{4}}+\frac{TS_{3}}{\rho _{3}}\right] $ \\ 
$A_{4E}\simeq L_{8}Z_{1}+L_{11}\frac{\rho _{3}}{\rho }\left( \frac{\rho _{1}%
}{\rho _{4}}Z_{1}-Z_{3}\right) $ \\ 
$A_{4P}\simeq L_{8}\left( \frac{v_{2}}{M_{2}}-\frac{v_{1}}{M_{1}}\right) -%
\frac{\rho _{3}\phi _{4}}{\rho \rho _{4}}L_{11}$ \\ 
$A_{4C}\simeq -\frac{R}{M_{1}C}L_{8}$%
\end{tabular}%
\end{equation}


\section{Discussion}
\label{sec:6}

\subsection{Nafion physicochemical properties}

\label{sec:61} In order to approximate these complex equations, we are going to estimate
the different terms. To do this, we focus on a particular electroactive
polymer, Nafion, and we restrict ourself to the isothermal case. \newline

\noindent The physicochemical properties of the dry polymer are well documented; its
molecular weight $M_{3}$ is between $\textit{10}^{\textit{2}}$ and 
$\textit{10}^{\textit{3}}\;kg\;mol^{-1}$ \cite{Heitner-Wirguin} and its mass 
density $\rho _{3}^{0}$ is close to $\textit{2.1}\;\textit{10}^{\textit{3}}%
\;kg\;m^{-3}$ \cite{nemat2000}. Its equivalent weight $M_{eq}$,
that is to say, the weight of polymer per mole of ionic sites is $%
\textit{1.1}\;kg\;eq^{-1}$ \cite{Gebel}. We deduce the electric charge per unit of
mass $Z_{3}=-\frac{F}{M_{eq}} =\textit{9}\;\textit{10}^{\textit{4}}\;C\;kg^{-1}$ where $%
F =\textit{96487}\;C\;mol^{-1}$ denotes the Faraday's constant. The cations may be $%
H^{+}$, $Li^{+}$ or $Na^{+}$ ions; we use an average molar mass $M_{1}\sim
\textit{10}^{\textit{-2}}\;kg\;mol^{-1}$, which corresponds to a mass electric charge $%
Z_{1}\sim \textit{10}^{\textit{7}}\;C\;kg^{-1}$. The cations partial molar volume $v_{1}$ is
on the order of $\frac{M_{1}}{\rho _{4}^{0}}\sim \textit{10}^{\textit{-5}}\;m^{3}\;mol^{-1}$.
The solvent molar mass $M_{2}$ is equal to $\textit{18}\;\textit{10}^{\textit{-3}}\;kg\;mol^{-1}$ and
its mass density $\rho _{2}^{0}$ \ is $\textit{10}^{\textit{3}}\;kg\;m^{-3}$; its partial
molar volume $v_{2}$ is approximately equal to $\textit{18}\;\textit{10}^{\textit{-6}}\;m^{3}\;mol^{-1}$%
, which is the molar volume of pure solvent. The dynamic viscosity of water $%
\eta _{2}$ is $\textit{10}^{\textit{-3}}\;Pa\;s$. \newline

\noindent When the polymer is saturated with water, the solution mass fraction is
usually between $\textit{20\%}$ and $\textit{25\%}$ if the counterion is a proton \cite%
{Cappadonia}. It corresponds to a volume fraction $\phi _{4}$ between $\textit{34\%}$
and $\textit{41\%}$. According to P. Choi \cite{Choi}, each anion is then surrounded
by an average of $14$ molecules of water, which corresponds to a porosity of 
$\textit{32\%}$. In the case of a counterion $Li^{+}$ or $Na^{+}$, S. Nemat-Nasser
and J. Yu Li \cite{nemat2000} indicate that the volume increases by $\textit{44.3\%}$
and $\textit{61.7\%}$ respectively between the dry and the saturated polymer, which
corresponds to porosities equal to $\textit{31\%}$ and $\textit{38\%}$. Thereafter we use an
average value $\phi _{4}\sim \textit{35\%}$. We deduce the mass densities of the
complete material, cations, solvent and solid relative to the volume of the
whole material%
\begin{equation}
\begin{tabular}{ll}
$\rho _{1}\sim 14\;kg\;m^{-3}\qquad \qquad $ & $\rho _{2}\sim
0.35\;10^{3}\;kg\;m^{-3}$ \\ 
$\rho _{3}\sim 1.4\;10^{3}\;kg\;m^{-3}\qquad \qquad $ & $\rho \sim
1.8\;10^{3}\;kg\;m^{-3}$%
\end{tabular}%
\end{equation}%
The cations molar fraction relative to the liquid phase and the anions molar
concentration, which is equal to the average cations concentration, can be
written 
\begin{equation}
x\sim 7\%\qquad \qquad C\sim 4\;10^{3}\;mol\;m^{-3}
\end{equation}%
In the following, we suppose that the temperature is $T=\textit{300}\;K$. Regarding to the
electric field, it is typically about $\textit{10}^{\textit{4}}\;V\;m^{-1}$ \cite{nemat2000}.

\subsection{Rheological equation}

\label{sec:62} We have shown that the rheological equation of the complete material is
identified with a Kelvin-Voigt model%
\begin{equation}
\boldsymbol{\sigma }=\lambda \left( tr\boldsymbol{\varepsilon }\right) \boldsymbol{1}+2G%
\boldsymbol{\varepsilon }+\lambda _{v}\left( tr\overset{\bullet }{\boldsymbol{\varepsilon }}%
\right) \boldsymbol{1}+2\mu _{v}\overset{\bullet }{\boldsymbol{\varepsilon }}
\end{equation}%
where $\lambda $ and $G$ are respectively the first Lam\'{e} constant and
the shear modulus of the whole material. $\lambda _{v}$ and $\mu _{v}$ are
viscoelastic coefficients%
\begin{equation}
\lambda _{v}\equiv \frac{L_{1}^{\prime }}{T}\qquad \qquad 2\mu _{v}\equiv 
\frac{L_{2}}{T}
\end{equation}%
Nafion is a thermoplastic semi-crystalline ionomer. M.\ N. Silberstein and
M.\ C. Boyce represent the polymer by a Zener model \cite{Silberstein2010}.
The elastic coefficients of the dry polymer can be deduced from their
measures%
\begin{equation}
G_{3}\sim 1.1\;10^{8}\;Pa\qquad \lambda _{3}\sim 2.6\;10^{8}\;Pa\qquad
E_{3}\sim 3\;10^{8}\;Pa\qquad \nu _{3}\sim 0.36
\end{equation}%
where $G_{3}$ is the shear modulus, $\lambda _{3}$ the first Lam\'{e}
constant, $E_{3}$ the Young's modulus and $\nu _{3}$ the Poisson's ratio of
the solid phase. Young's modulus is in good agreement with the values cited
in \cite{Satterfield2009}. They also correspond to the typical values of this
kind of polymer, especially Poisson's ratio, which is usually close to $\textit{0.33}$
below the glass transition temperature and to $\textit{0.5}$ around the transition
temperature \cite{ferry}. \newline

\noindent When the polymer is saturated with water, the elastic coefficients vary;
water has a plasticising effect \cite{Kundu,Bauer}. We obtain the
following values \cite{Silberstein2010,Satterfield2009,Bauer}%
, which are in agreement with the usual ones \cite{ferry}%
\begin{equation}
G\sim 4.5\;10^{7}\;Pa\qquad \lambda \sim 3\;10^{8}\;Pa\qquad E\sim
1.3\;10^{8}\;Pa\qquad \nu \sim 0.435
\end{equation}
Viscoelastic coefficients can be deduced from uniaxial tension tests \cite
{Silberstein2010,Satterfield2009,Silberstein2011}

\begin{equation}
E_{v}=\frac{\mu _{v}\left(3\lambda _{v}+2\mu _{v} \right)}{\lambda _{v}+\mu _{v}}%
\sim 1.2\;10^{8}\;Pa\;s
\end{equation}

\noindent
The viscoelastic coefficients $\lambda _{v}$ and $\mu _{v}$ (or $E_v$) can be estimated from the 
relaxation times according to traction and shear tests. Typically, the relaxation time for a traction is of the order 
$\theta_E \sim \textit{15} \; \textit{s}$ for the saturated Nafion polymer   \cite{Silberstein2010,Silberstein2011,Silberstein2008}. 
The shear relaxation time is usually of the same order of the traction one : $\theta_{\mu} \sim \theta_E$   \cite{ferry,Strobl,Combette}. The 
viscoelastic coefficients are given by the relations $E_v = E \theta_{E}$ and $\mu_v = G \theta_{\mu}$ for the 
traction and shear viscoelastic modulus, respectively. Therefore, the phenomenological coefficients are given by 
\begin{equation}
\lambda \sim \;3\; 10^8\; Pa \qquad G \sim 4.5\;10^{7}\;Pa \qquad
\lambda_v \sim \; 7 \; 10^8\; Pa \: s  \qquad \mu_v \sim 10^{8}\;Pa \: s 
\end{equation}

\noindent
Accordingly, we deduce from (84) 

\begin{equation}
L_{1}^{^{\prime }} \sim 2.1 \; 10^{11} \;Pa \: s  \: K \qquad \qquad L_{2}\sim 6  \; 10^{10} \;Pa \: s  \: K 
\end{equation}%

\noindent
It is worthwhile noting that these viscoelastic phenomenological coefficients depend very strongly on the solvent concentration and on the temperature, especially if the operating temperature of the polymer is close to that of the glass transition. 
In addition, the molecular relaxation time is of the order of $\textit{10} \; \textit{s}$ just bellow the glass transition \cite{Strobl,Combette}.

\subsection{Nernst-Planck equation}

\label{sec:63} In the following, we focus on the isothermal case. Considering the previous
numerical estimations, we can write in a first approximation%
\begin{equation}
Z_{1}>>Z_{3}\qquad \qquad \rho \sim \rho _{2}\sim \rho _{3}>>\rho _{1}\qquad
\qquad \rho _{1}Z_{1}\sim \rho _{4}Z_{3}
\end{equation}%
Moreover, the non-diagonal phenomenological coefficients are usually small
compared to the diagonal ones; we suppose that%
\begin{equation}
L_{3}\gtrsim L_{4},L_{5}\qquad \qquad L_{7}\gtrsim L_{4},L_{8}\qquad \qquad
L_{11}\gtrsim L_{5},L_{8}
\end{equation}%
One deduces%
\begin{equation}
\mathbf{J_{1}}\simeq \frac{L_{7}Z_{1}}{T}\mathbf{E}+\frac{1}{%
T}\left[ \frac{v_{2}}{M_{2}}\left( L_{7}-\frac{\rho _{3} L_{8}}{\rho }\right)
-\frac{v_{1} L_{7}}{M_{1}}\right] \boldsymbol{\nabla} p-\frac{R L_{7}}{M_{1}C}%
\boldsymbol{\nabla} C
\end{equation}%
that is to say%
\begin{equation}
\begin{tabular}{l}
$\mathbf{V_{1}}\simeq -\frac{RL_{7}}{M_{1}\rho _{1}C} %
\left\{ \boldsymbol{\nabla} C-\frac{M_{1}CZ_{1}}{RT}\mathbf{E} \right.$ \\ 
$\qquad \qquad \left. +\frac{Cv_{1}}{RT}\left[ 1-\frac{M_{1}v_{2}}{M_{2}v_{1}}\left( %
1-\frac{\rho _{3} L^{8}}{\rho L^{7}} \right) \right] \boldsymbol{\nabla} p %
\right\} + \mathbf{V_{2}}$%
\end{tabular}%
\end{equation}%
This expression is identified with the Nernst-Planck equation \cite{Lakshmi,Schlogl,Schlogl2}%
\begin{equation}
\mathbf{V_{1}}=-\frac{D}{C}\left[ \boldsymbol{\nabla} C-\frac{%
Z_{1}M_{1}C}{RT}\mathbf{E}+\frac{Cv_{1}}{RT}\left( 1-\frac{M_{1}}{%
M_{2}}\frac{v_{2}}{v_{1}}\right) \boldsymbol{\nabla} p\right] +%
\mathbf{V_{2}}
\end{equation}%
where $D$ denotes the mass diffusion coefficient of the cations in the
liquid phase and $v_{1}$ their partial molar volume. This equation expresses
the equilibrium of an ions mole under the action of four forces : the Stokes
friction force $-6\pi \eta _{2}aN_{a}\left( \mathbf{V_{1}}-%
\mathbf{V_{2}}\right) $, the pressure force $-v_{1}\left( 1-\frac{%
M_{1}}{M_{2}}\frac{v_{2}}{v_{1}}\right) \boldsymbol{\nabla} p$, the
electric force $Z_{1}M_{1}\mathbf{E}$ and the thermodynamic force $%
-M_{1}\boldsymbol{\nabla} \mu _{1}$; $N_{a}$ denotes the Avogadro
constant and $a$ the ion hydrodynamic radius, i.e. the radius of the
hydrated ion. The proton mass diffusion coefficient $D=\frac{RT}{6\pi \eta
_{2}aN_{a}}$ is about $\textit{2}\;\textit{10}^{\textit{-9}}$ $m^{2}s^{-1}$ 
\cite{Zawodsinski,Kreuer2001}. The proportionality factor $1-\frac{M_{1}} %
{M_{2}}\frac{v_{2}}{v_{1}}$ reduces the mass pressure force exerted on the solution to the
cations; it is therefore of the order of $x$. We obtain by identification%
\begin{equation}
\begin{tabular}{l}
 
$L_{8}<<L^{7}$
\end{tabular}%
\end{equation}
We can now estimate the order of magnitude of the different terms of this
equation. The concentration gradient $\left\vert \boldsymbol{\nabla} %
C\right\vert $ can be evaluated by dividing the average concentration of
anions (or cations) by the polymer film thickness. This thickness $e$ is
typically about $\textit{200}\;\mu m$ \cite{nemat2000}, which provides a
concentration gradient of the order of $\textit{2}\;\textit{10}^{\textit{7}}\;mol\;m^{-4}$. 
More precisely, numerical studies show that cations gather near the electrode of
opposite sign. The concentration gradient is thus higher in certain zones
than the previous evaluation. These simulations enable to estimate the
maximal concentration gradient at $\textit{7}\;\textit{10}^{\textit{8}}\;mol\;m^{-4}$ 
\cite{nemat2002} or at $\textit{10}^{\textit{7}}\;mol\;m^{-4}$ \cite{Farinholt}. Thence%
\begin{equation}
\left\vert \boldsymbol{\nabla} C\right\vert \precsim 10^{8}\;mol\;m^{-4}
\end{equation}
The pressure gradient can be roughly estimated by dividing the air pressure
by the strip thickness, which provides a value about $\textit{5}\;\textit{10}^{\textit{8}}\;Pa\;m^{-1}$, %
or using the Darcy's law; the average fluid velocity can be estimated from the response %
time of the polymer strip $\tau \sim \textit{1}$ to $\textit{10}\;s$ \cite{nemat2000}%
\begin{equation}
\left\vert \mathbf{V_{4,moy}}\right\vert \sim \frac{e}{\tau }\sim
10^{-4}\;m\;s^{-1}
\end{equation}%
Otherwise, the characteristic size $d$ of the hydrated polymer pores is
about $\textit{100}\;\textit{\AA}$ \cite{Gebel,Pineri}. We can deduce the polymer
intrinsic permeability $K$, which is on the order of the square of the pore
size ($\textit{10}^{\textit{-16}}\;m^{2}$). Darcy's law then provides%
\begin{equation}
\left\vert \boldsymbol{\nabla} p\right\vert \sim \frac{\eta _{2}}{K}%
\left\vert \mathbf{V_{4,moy}}\right\vert \sim 10^{9}\;Pa\;m^{-1}
\end{equation}%
This is in good agreement with the previous estimation. \newline

\noindent We finally obtain the following orders of magnitude for the different terms
of the Nernst-Planck equation%
\begin{equation}
\begin{tabular}{ll}
$\left\vert \boldsymbol{\nabla} C\right\vert $ & $\lesssim
10^{8}\;mol\;m^{-4}$ \\ 
$\frac{M_{1}C}{RT}Z_{1}\left\vert \mathbf{E}\right\vert $ & $\sim
1.6\;10^{9}\;mol\;m^{-4}$ \\ 
$\frac{Cv_{1}}{RT}\left( 1-\frac{M_{1}}{M_{2}}\frac{v_{2}}{v_{1}}\right)
\left\vert \boldsymbol{\nabla} p\right\vert $ & $\sim
1.1\;10^{3}\;mol\;m^{-4}$%
\end{tabular}%
\end{equation}%
Cations mainly move under the actions of the electric field and the mass
diffusion; pressure gradient effect is negligible.

\subsection{Generalized Darcy's law}

\label{sec:64} In the isothermal case, the mass diffusion flux of the solution in the solid
can be approximated 
\begin{equation}
\begin{tabular}{l}
$\mathbf{J_{4}}\simeq \frac{1}{T}\left[ L_{8}Z_{1}+L_{11}\frac{\rho
_{3}}{\rho }\left( \frac{\rho _{1}}{\rho _{4}}Z_{1}-Z_{3}\right) \right] 
\mathbf{E}-\frac{R}{M_{1}C}L_{8}\boldsymbol{\nabla} C$ \\ 
$\quad +\frac{1}{T}\left[ L_{8}\left( \frac{v_{2}}{M_{2}}-\frac{v_{1}}{M_{1}}%
\right) -\frac{\rho _{3}\phi _{4}}{\rho \rho _{4}}L_{11}\right] 
\boldsymbol{\nabla} p$%
\end{tabular}%
\end{equation}%
The pressure term must be identified with Darcy's law%
\begin{equation}
\frac{1}{T\rho _{4}}\left[ L_{8}\left( \frac{v_{1}}{M_{1}}-\frac{v_{2}}{M_{2}%
}\right) +\frac{\rho _{3}\phi _{4}}{\rho \rho _{4}}L_{11}\right] \sim \frac{K%
}{\eta _{2}\phi _{4}}
\end{equation}%
where $K$ denotes the intrinsic permeability of the solid phase. Considering
the previous estimation of $L_{8}$, the first term is negligible, then we can
compute again $L_{11}$%
\begin{equation}
L_{11}\sim \frac{KT}{\eta _{2}\phi _{4}^{2}}\frac{\rho _{2}^{2}\rho }{\rho
_{3}}\sim 3.8\;10^{-5}\;kg\;s\;K\;m^{-3}>>L^{8}
\end{equation}%
The constitutive equation becomes%
\begin{equation}
\mathbf{V_{4}}-\mathbf{V_{3}}\simeq -\frac{K}{\eta _{2}\phi
_{4}}\left[ \boldsymbol{\nabla} p-\rho _{2}^{0}\left( \frac{\rho _{1}}{\rho
_{4}}Z_{1}-Z_{3}\right) \mathbf{E}\right] -\frac{R}{M_{1}C\rho _{4}}%
L^{8}\boldsymbol{\nabla} C
\end{equation}%
The orders of magnitude of the different terms are%
\begin{equation}
\begin{tabular}{ll}
$\frac{K}{\eta _{2}\phi _{4}}\left\vert \boldsymbol{\nabla} p\right\vert $
& $\sim 2.8\;10^{-4}\;m\;s^{-1}$ \\ 
$\frac{K\rho _{2}^{0}}{\eta _{2}\phi _{4}}\left( \frac{\rho _{1}}{\rho _{4}}%
Z_{1}-Z_{3}\right) \left\vert \mathbf{E}\right\vert $ & $\sim
1.1\;m\;s^{-1}$ \\ 
$\frac{R}{M_{1}C\rho _{4}}L^{8}\left\vert \boldsymbol{\nabla} C\right\vert $
& $<<2\;10^{-6}\;m\;s^{-1}$%
\end{tabular}%
\end{equation}%
The phenomenological equation thus obtained can be identified at a first
approximation with a generalized Darcy's law%
\begin{equation}
\mathbf{V_{4}}-\mathbf{V_{3}}\simeq -\frac{K}{\eta _{2}\phi
_{4}}\left[ \boldsymbol{\nabla} p-\rho _{4}^{0}\left( Z_{4}-Z_{3}\right) 
\mathbf{E}\right] 
\end{equation}%
In this expression, $\frac{1}{\rho _{4}^{0}}\boldsymbol{\nabla} p$
represents the mass pressure force and $\left( Z_{4}-Z_{3}\right) 
\mathbf{E}$ is the mass electric force. The second term expresses
the motion of the solution under the action of the electric field; it
consists in an electroosmotic term. \newline

\noindent When an electric field is applied, the
cations distribution becomes very heterogeneous \cite{nemat2002,Farinholt}. 
Three regions can be distinguished
\begin{itemize}
\item Around the negative electrode, where cations gather, $Z_{4}>>Z_{3}$
and 
\begin{equation*}
\mathbf{V_{4}}-\mathbf{V_{3}}=\frac{K}{\eta _{2}\phi _{4}}%
\left( \rho _{4}^{0}Z_{4}\mathbf{E}-\boldsymbol{\nabla} p\right)
\end{equation*}%
The electric force exerted on the solution is due to the cations charge; we
find out the expression obtained by M.A. Biot \cite{Biot}.

\item Near the positive electrode, where the cation concentration is very
low, $Z_{4}<<Z_{3}$ and%
\begin{equation}
\mathbf{V_{4}}-\mathbf{V_{3}}=-\frac{K}{\eta _{2}\phi _{4}}%
\left( \rho _{4}^{0}Z_{3}\mathbf{E}+\boldsymbol{\nabla} p\right)
\end{equation}%
$\rho _{4}^{0}Z_{3}\mathbf{E}$ represents the electric force exerted
on the anions relative to the volume of the solution. This result
corresponds to the expression obtained by Grimshaw et al \cite{nemat2000,Grimshaw}. 
The solution motion is due to the attractive force exerted
on the cations by the solid.

\item In the center of the strip, $Z_{4}\sim Z_{3}$. The solution electric
charge is partially balanced with the solid one, and the mass electric force
exerted on the solution is proportional to the net charge $\left(
Z_{4}-Z_{3}\right) $.
\end{itemize}


\section{Conclusion}

\label{sec:concl} We have studied an ionic electro-active polymer. When this electrolyte is
saturated with water, it is fully dissociated and releases cations of small
size, while anions remain bound to the polymer backbone. We have depicted
this system as the superposition of three systems : a solid component, the
polymer backbone negatively charged, which is assimilated to a deformable
porous medium; and an ionic liquid solution, composed by the free cations
and the solvent (the water); these three components move with different
velocity fields. In a previous article \cite{Tixier}, we have established \ the
conservation laws of the two phases : mass continuity equation, Maxwell's
equations, linear momentum conservation law and energy balance laws.
Averaging these equations over the R.V.E. and using the material derivative
concept, we obtained the conservation laws of the complete material.\newline

\noindent In this paper, we derive the entropy balance law and the thermodynamic
relations using the same method. We deduce the entropy production and
indentify the generalized forces and fluxes. Then we can write the
constitutive equations of the complete material. The first one links the
stress tensor with the strain tensor; the saturated polymer satisfies a
Kelvin-Voigt model. The three others are vectorial equations, including a
generalized Fourier's law. Focusing on the isothermal case, we also obtain a
generalized Darcy's law and find out the Nernst-Planck equation. Using the
Nafion physico-chemical properties, we estimate the phenomenological
coefficients. This enables an evaluation of the different terms of the
equations. \newline

\noindent We now plan to compare these results with experimental data published in the
literature. This should allow us to improve our model. Other possibility of
improvement of the model should consider the Zener model for the
viscoelastic behavior of the polymer.


\section{Appendix A : Particle derivatives and material derivative}

\label{sec:App1} In order to write the balance equations of the whole material, we use the
material derivative $\frac{D}{Dt}$ defined in our previous paper \cite%
{Tixier}.

\noindent Indeed, the different phases do not move with the same velocity : velocities
of the solid and the solution are a priori different. For a quantity $g$, we
can define particle derivatives following the motion of the solid $(\frac{%
d_{3}}{dt})$, the solution $(\frac{d_{4}}{dt})$ or the interface $(\frac{%
d_{i}}{dt})$ as
\begin{equation}
\frac{d_{k}g}{dt}=\frac{\partial g}{\partial t}+\boldsymbol{\nabla} g\cdot 
\mathbf{V_{k}}
\end{equation}%
Let us consider an extensive quantity of density $g\left( \mathbf{r}%
,t\right) $ relative to the whole material.%
\begin{equation}
g=g_{3}+g_{4}+g_{i}
\end{equation}%
where $g_{3}$, $g_{4}$ and $g_{i}$ are the densities relative to the total
actual volume attached to the solid, the solution and the interface,
respectively. Material derivative enables to calculate the variation of $%
g\left( \mathbf{r},t\right) $ following the motion of the different
phases \cite{Coussy95,Biot77,Coussy89}%
\begin{equation}
\rho \frac{D}{Dt}\left( \frac{g}{\rho }\right) =\sum\limits_{k=3,4,i}\rho
_{k}\frac{d_{k}} {dt} \left( \frac{g_{k}}{\rho _{k}}\right)%
=\sum\limits_{k=3,4,i}\frac{\partial g_{k}}{\partial t}+div\left( g_{k}%
\mathbf{V_{k}}\right)
\end{equation}%
This derivative must not be confused with the derivative $\frac{d}{dt}$
following the barycentric velocity $\mathbf{V}$.


\section{Appendix B : Interface modelling}

\label{sec:App2} In practice, contact area between phases $3$ and $4$ has a certain
thickness; extensive physical quantities vary from one bulk phase to the
other one. This complicated reality can be modelled by two uniform bulk
phases separated by a discontinuity surface $\Sigma $ whose localization is
arbitrary. Let $\Omega $ be a cylinder crossing $\Sigma $, whose bases are
parallel to $\Sigma $. We denote by $\Omega _{3}$ and $\Omega _{4}$ the
parts of $\Omega $ respectively included in phases $3$ and $4$.

\noindent The continuous quantities relative to the contact zone are identified by a
superscript $^{0}$ and no subscript. The microscale surface entropy $%
S_{i}^{0}$ and the microscale surface entropy production $s_{i}^{0}$ are
defined by%
\begin{equation}
\begin{tabular}{l}
$S_{i}^{0}=\underset{\Sigma \longrightarrow 0}{\lim }\frac{1}{\Sigma }%
\left\{ \int_{\Omega }S^{0}dv-\int_{\Omega _{3}}S_{3}^{0}dv-\int_{\Omega
_{4}}S_{4}^{0}dv\right\} $ \\ 
$s_{i}^{0}=\underset{\Sigma \longrightarrow 0}{\lim }\frac{1}{\Sigma }%
\left\{ \int_{\Omega }s^{0}dv-\int_{\Omega _{3}}s_{3}^{0}dv-\int_{\Omega
_{4}}s_{4}^{0}dv\right\} $%
\end{tabular}%
\end{equation}%
where $\Omega _{3}$ and $\Omega _{4}$ are small enough so that $S_{3}^{0}$, $%
S_{4}^{0}$, $s_{3}^{0}$ and $s_{4}^{0}$ are constant. Their averages over
the R.V.E. are the volume quantity $S_{i}$ and $s_{i}$%
\begin{equation}
S_{i}=\left\langle \chi _{i}S_{i}^{0}\right\rangle \qquad \qquad
s_{i}=\left\langle \chi _{i}s_{i}^{0}\right\rangle
\end{equation}%
We arbitrarily fix the interface position in such a way that it has no mass
density
\begin{equation}
\rho _{i}^{0}=\lim\limits_{\Sigma \longrightarrow 0}\frac{1}{\Sigma }\left\{
\int_{\Omega }\rho ^{0}dv-\int_{\Omega _{3}}\rho _{3}^{0}dv-\int_{\Omega
_{4}}\rho _{4}^{0}dv\right\} =0  \label{DefI}
\end{equation}%

\noindent Neglecting the heat flux along the interfaces, the balance equation of the
interfacial quantity $S_{i}^{0}$ is written as \cite{Ishii06}%
\begin{equation}
\frac{\partial S_{i}^{0}}{\partial t}+div_{s}\left( S_{i}^{0}\mathbf{%
V_{i}^{0}}\right) =\sum\limits_{3,4}\left[ S_{k}^{0}\left( \mathbf{%
V_{k}^{0}}-\mathbf{V_{i}^{0}}\right) .\mathbf{n_{k}}+%
\mathbf{\Sigma _{k}^{0}}.\mathbf{n_{k}}\right] +s_{i}^{0}
\end{equation}%
where $div_{s}$ denotes the surface divergence operator. Averaging this
equation over the R.V.E. provides%
\begin{equation}
\frac{\partial S_{i}}{\partial t}+div\left( S_{i}\mathbf{V_{i}}%
\right) =\sum\limits_{3,4}\left\langle \chi _{i}\mathbf{\Sigma
_{k}^{0}}.\mathbf{n_{k}}\right\rangle +s_{i}  \label{Si}
\end{equation}

\noindent Interfacial Gibbs equation derives from the entropy balance equation (\ref%
{Si}) and from the internal energy balance equation established in \cite%
{Tixier}%
\begin{equation}
\frac{d_{i}U_{i}}{dt}=T_{i}\frac{d_{i}S_{i}}{dt}
\end{equation}%
remarking that entropy production $s_{i}$ and diffusion current $%
\mathbf{i_{i}}$ \ cancel at equilibrium. The interface has no mass
density; as a result, there is no mass exchange term in this relation.

\noindent In the same way, Euler's relation and Gibbs-Duhem relation write%
\begin{equation}
U_{i}-T_{i}S_{i}=0
\end{equation}%
\begin{equation}
S_{i}\frac{d_{i}T_{i}}{dt}=0
\end{equation}


\section{Appendix C : Small deformation hypothesis}

\label{sec:App3} In the case of small deformations, the Green-Lagrange finite strain tensor
come down to the Cauchy's infinitesimal strain tensor $\boldsymbol{\varepsilon _{3}^{0}}$%
\begin{equation}
\boldsymbol{\varepsilon _{3}^{0}}=\frac{1}{2}\left(\boldsymbol{\nabla} \mathbf{u_{3}^{0}}+%
\boldsymbol{\nabla} \mathbf{u_{3}^{0}}^{T}\right)
\end{equation}%
where $\mathbf{u_{3}^{0}}$ is the displacement vector \cite{Coussy95}%
. The solid phase velocity is defined by%
\begin{equation}
\mathbf{V_{3}^{0}}=\frac{d_{3}^{0}}{dt}\left( \mathbf{%
u_{3}^{0}}\right)
\end{equation}%
The small deformation hypothesis results in%
\begin{equation}
\left\vert \boldsymbol{\nabla} \mathbf{u_{3}^{0}} \right\vert <<1\qquad and%
\qquad \left\vert \boldsymbol{\nabla} \mathbf{V_{3}^{0}}\right\vert <<1
\end{equation}

\noindent Let $\mathbf{A}$, a vectorial quantity. The particles derivative of $%
\mathbf{A}$ following the motion of the solid phase identifies with%
\begin{equation}
\frac{d_{3}^{0}}{dt}\left( \mathbf{A}\right) \equiv \frac{\partial }{%
\partial t}\left( \mathbf{A}\right) +\boldsymbol{\nabla} \left( \mathbf{A}\right) .\mathbf{V_{3}^{0}}
\end{equation}%
Small deformation assumption leads to%
\begin{equation}
\frac{d_{3}^{0}}{dt}\left[ div\left( \mathbf{A}\right) \right]
\simeq div\left( \frac{d_{3}^{0}\mathbf{A}}{dt}\right)
\end{equation}%
\begin{equation}
\frac{d_{3}^{0}}{dt}\left[ \boldsymbol{\nabla}  \mathbf{A}\right] %
\simeq \boldsymbol{\nabla}\left( \frac{d_{3}^{0}\mathbf{A}}{dt}\right)
\end{equation}%
One deduces%
\begin{equation}
\frac{d_{3}}{dt}\boldsymbol{\varepsilon _{3}}\simeq \frac{d_{3}^{0}}{dt}\boldsymbol{\varepsilon _{3}^{0}}
\end{equation}


\section{Notations}

\label{sec:not} $k=1,2,3,4,i$ subscripts respectively represent cations, solvent, solid,
solution (water and cations) and interface; quantities without subscript
refer to the whole material. Superscript $^{0}$ denotes a local quantity;
the lack of superscript indicates average quantity at the macroscopic\
scale. Microscale volume quantities are relative to the volume of the phase,
average quantities to the volume of the whole material. Superscripts $^{s}$
and $^{a}$ respectively indicate the deviatoric and the antisymmetric parts
of a second-rank tensor, and $^{T}$ its transpose.

\begin{description}
\item $C$ : cations molar concentration (relative to the liquid phase);

\item $D$ : mass diffusion coefficient of the cations in the liquid phase;

\item $\mathbf{D}$ : electric displacement field;

\item $E$, $E_{3}$ : Young's modulus;

\item $\mathbf{E}$ : electric field;

\item $E_{c}$, $E_{c\Sigma }$ : kinetic energy density;

\item $F=96487\;C\;mol^{-1}$ : Faraday's constant ;

\item $G$, $G_{3}$ : shear modulus;

\item $\mathbf{I}$ : current density vector;

\item $\mathbf{i}\ $($\mathbf{i^{\prime }}$, $%
\mathbf{i_{k}}$, $\mathbf{i_{k}^{0}}$) : diffusion current;

\item $\mathbf{J_{k}}$ : mass diffusion flux;

\item $K$ : intrinsic permeability of the solid phase;

\item $L_{i},L_{i}^{\prime }$ : phenomenological coefficients;

\item $M_{k}$ : molar mass of component $k$;

\item $M_{eq}$ : equivalent weight (weight of polymer per mole of sulfonate
groups);

\item $\mathbf{n_{k}}$ : outward-pointing unit normal of phase $k$;

\item $p$ ($p_{k}$, $p_{k}^{0}$) : pressure;

\item $\mathbf{Q}$ ($\mathbf{Q^{\prime }}$, $\mathbf{%
Q_{k}^{0}}$) : heat flux;

\item $R=8,314\;J\;K^{-1}$ : gaz constant;

\item $s$ ($s_{k}^{0}$, $s_{k}$) : rate of entropy production;

\item $S$ ($S_{k}^{0}$, $S_{k}$) : entropy density;

\item $T$ ($T_{k}$, $T_{k}^{0}$) : absolute temperature;

\item $U$ ($U_{\Sigma }$,$U_{k}$, $U_{k}^{0}$) : internal energy density;

\item $\mathbf{u}$ ($\mathbf{u_{3}^{0}}$, $\mathbf{%
u_{3}}$) : displacement vector;

\item $v_{k}$ : partial molar volume of component $k$ (relative to the
liquid phase);

\item $\mathbf{V}$ ($\mathbf{V_{k}}$, $\mathbf{%
V_{k}^{0}}$) : velocity;

\item $x$ : cations mole fraction (relative to the liquid phase);

\item $Z$ ($Z_{k}$, $Z_{k}^{0}$) : total electric charge per unit of mass;

\item $\varepsilon $ ($\varepsilon _{k}^{0}$) : permittivity;

\item $\boldsymbol{\varepsilon }$ ($\boldsymbol{\varepsilon _{k}}$, $\boldsymbol{%
\varepsilon _{k}^{0}}$) : strain tensor;

\item $\eta _{2}$ : dynamic viscosity of water;

\item $\lambda $, $\lambda _{3}$ : first Lam\'{e} constant;

\item $\lambda _{v}$\textit{, }$\mu _{v}$\textit{, }$E_{v}$\textit{\ }: viscoelastic
coefficients;

\item $\nu $, $\nu _{3}$ : Poisson's ratio;

\item $\mu _{k}$, $\mu _{k}^{0}$ ($\mu _{k,mol}^{0}$) : mass (molar)
chemical potential;

\item $\rho $ ($\rho _{k}$, $\rho _{k}^{\prime }$, $\rho _{k}^{0}$) : mass
density;

\item $\boldsymbol{\sigma }$ ($\boldsymbol{\sigma _{k}}$) : stress tensor;

\item $\boldsymbol{\sigma ^{v}}$ : dynamic stress
tensor;

\item $\boldsymbol{\sigma ^{e}}$ ($\boldsymbol{\sigma _{k}^{e}},\boldsymbol{\sigma
_{k}^{0e}}$) : equilibrium stress tensor;

\item $\mathbf{\Sigma }$ ($\mathbf{\Sigma ^{\prime }}$, $%
\mathbf{\Sigma _{k}^{0}}$, $\mathbf{\Sigma _{k}}$) : entropy
flux vector;

\item $\phi _{k}$ : volume fraction of phase $k$;

\item $\chi _{k}$ : function of presence of phase $k$ ;
\end{description}


\end{document}